\documentclass[10pt]{article}

\usepackage{amsmath}
\usepackage{amssymb}
\usepackage{graphicx}

\title{\textbf{Foliable Operational Structures for General Probabilistic Theories}}

\author{Lucien Hardy\\
\textit{Perimeter Institute,}\\
\textit{31 Caroline Street North,}\\
\textit{Waterloo, Ontario N2L 2Y5, Canada}}

\begin{document}

\maketitle

\begin{abstract}
In this chapter a general mathematical framework for probabilistic theories of 
operationally understood circuits is laid out. Circuits are comprised of 
operations and wires. An operation is one use of an apparatus and a wire is a 
diagrammatic device for showing how apertures on the apparatuses are placed
next to each other.  Mathematical objects are defined in terms of the circuit
understood graphically.  In particular, we do not think of the circuit as 
sitting in a background time.  Circuits can be foliated by hypersurfaces 
comprised of sets of wires.  Systems are defined to be associated with wires. A 
closable set of operations is defined to be one for which the probability 
associated with any circuit built from this set is independent both of choices 
on other circuits and of extra circuitry that may be added to outputs from this 
circuit.  States can be associated with circuit fragments corresponding to 
preparations. These states evolve on passing through circuit fragments 
corresponding to transformations.  The composition of transformations is 
treated.  A number of theorems are proven including one which rules out 
quaternionic quantum theory.   The case of locally tomographic theories (where 
local measurements on a systems components suffice to determine the global 
state) is considered.  For such theories the probability can be calculated for 
a circuit from matrices pertaining the operations that comprise that circuit.  
Classical probability theory and quantum theory are exhibited as examples in this framework.
\end{abstract}

\section{Introduction}

Prior to Einstein's 1905 paper \cite{Einstein1905} laying the foundations of special relativity it was known that Maxwell's equations are invariant under the Lorentz transformations. Mathematically the Lorentz transformations are rather complicated and it must have been unclear why nature would choose these transformations over the rather more natural looking Galilean transformations.  Further, there was a understanding of the physical reasons for Galilean transformations in terms of boosts and the additive nature of velocities.  We find ourselves in a similar situation today with respect to quantum theory.  Regarded as a probabilistic theory, it is much more complicated from a mathematical point of view than the rather natural equations of classical probability theory. And further, we can motivate classical probability by ordinary reasoning by imagining that the probabilities pertain to some underlying mutually exclusive set of possibilities.  The situation with respect to the Lorentz transformations was resolved by Einstein when he showed that they follow from two very reasonable postulates:  that the laws of physics are the same in all reference frames and that the speed of light in vacuum is independent of the motion of the source.  Once we see Einstein's reconstruction of the Lorentz transformations we have a sense that we understand why, at a fundamental level, nature prefers these over the mathematically simpler Galilean transformations.  We need something similar for quantum theory \cite{Fuchslittlemore}.

The subject of reconstructing quantum theory has seen something of a revival in the last decade \cite{RQTpirsa}.    Generally, to reconstruct quantum theory we write down a set of basic axioms or postulates which are supposed to be well motivated. They should not appear unduely mathematical.  Then we apply these in the context of some framework for physical theories and show that we obtain quantum theory.  This framework itself should be well motivated and may even follow from one or more of the given postulates.

\begin{figure*}
\centering
{\includegraphics[width=0.6\textwidth]{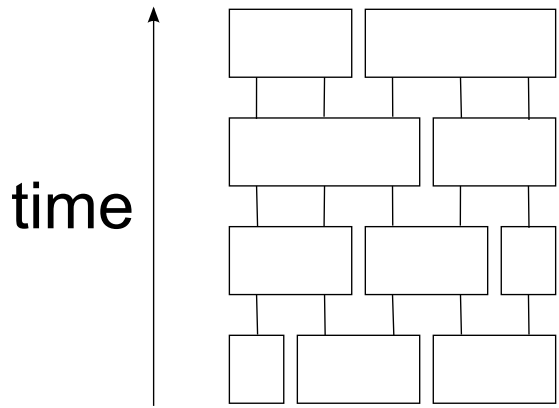}} \caption{\small A naive picture of operationalism. Systems pass through boxes with respect to a background time.}\label{naivepic}
\end{figure*}

The purpose of this chapter is to set up one such framework. This will be a framework for general probabilistic operational theories.  There is a large literature on this (see Section \ref{relatedwork}).  To construct the mathematics of such a framework we must first specify what we mean by our operational structure.  Only then can we add probability.  The mathematics associated with the part of this where we add probability has become fairly sophisticated.  However, a fairly simple minded point of view is usually taken with respect to setting up the operational structure upon which the whole endeavor is founded.  The picture normally adopted is of a system passing sequentially through various boxes representing operations or, more generally, of many systems where, at any given time, each system passes through a box with, possibly, the same box acting on more than one system at once (see Fig. \ref{naivepic}).  This simple picture is problematic for various reasons.  First, there is no reason why the types and number of systems going into a box be equal to the types and number emerging.  Second, the notion of system itself is not fully operational.  Third (and most significantly), this circuit is understood as being embedded in a background Newtonian time and this constitutes structure in addition to a purely graphical interpretation of the diagram (it matters how high on the page the box is placed since this corresponds to the time at which the operation happens).

To deal with these three points we set up a more general framework where (1) we allow  the number and types of systems going into a box to  be different from the the number and types of systems going out (2) give an operational definition of the notion of system (3) define our temporal concepts entirely in terms of the graphical information in the diagram.  This third point gives rise to a natural notion of spacelike hypersurface in such a way that multiple foliations are possible. Hence, we call this a {\it foliable operational structure}.

It is worth being careful to formulate the operational structure well since such structures form a foundation for general probabilistic theories.  Different operational structures can lead to different probabilistic frameworks.  Once we have an operational structure, we can introduce probabilities.  We then proceed along a fairly clear path introducing the notions of preparation, transformation, measurement, and associating mathematical objects with these that allow probabilities to be calculated. This gives a example of how an operational framework can be a foundation for a general probabilistic theory.  The foliable framework presented here is sufficient for the formulation of classical probability theory, quantum theory and potentially many theories beyond.

However, the foliable operational structure still, necessarily, has a notion of definite causal structure - when a system passes between two boxes that corresponds to a timelike separation (or null in the case of photons).  We anticipate that a theory of quantum gravity will be a probabilistic theory with indefinite causal structure.   If this is true then we need a more general framework than the one presented here for quantum gravity.  Preliminary ideas along this line have been presented in \cite{Hardy2}.  In future work it will be shown how the approach taken in this chapter can be generalized to theories without definite causal structure - that is non-foliable theories.  First we must specify a sufficiently general non-foliable operational structure and then add probabilities (see \cite{Hardypirsa1} for an outline of these ideas).

\section{Related work}\label{relatedwork}

The work presented here is a continuation of work initiated by the author in \cite{Hardy1} in which a general probabilistic framework, sometimes called the ${\bf r}$-${\bf p}$ framework (because these vectors represent effects and states), was obtained for the purpose of reconstructing quantum theory from some simple postulates.  In \cite{Hardy2,Hardy3} the author adapted the ${\bf r}$-${\bf p}$ for the purpose of describing a situation with indefinite causal structure to obtain a general probabilistic framework that might be suitable for a theory of quantum gravity.  The idea that states should be represented by joint rather than conditional probabilities used in these papers is also adopted here.  Preliminary versions of the work presented here can be seen in \cite{Hardypirsa1,Hardypirsa2}.

In this work we consider arbitrary foliations of circuits.  Thus we take a more space-time based approach. There are many other space-time based approach in the context of a discrete setting in the literature (particularly work on quantum gravity).  Sorkin builds a discrete model of space-time based on causal sets \cite{causalset}. Work in the consistent (or decoherent) histories tradition \cite{ch1, ch2, ch3, ch4} takes whole histories as the basic objects of study. A particularly important and influential piece of work is the quantum causal histories approach developed by Markopoulou \cite{Markopoulou}.  In this, completely positive maps are associated with the edges of a graph with Hilbert spaces living on the vertices.  Blute, Ivanov, and Panangaden \cite{Prakash} (see also \cite{Prakashpirsa}), motivated by Markopoulou's work, took the dual point of view with systems living on the edges (wires) and completely positive maps on the vertices.  The work of Blute {\it et al.}, though restricted to quantum theory rather than general probabilistic theories, bares much similarity with the present work.  In particular, similar notions of foliating circuits are to be found in that paper.  Leifer \cite{Leifer} has also done interesting work concerning the evolution of quantum systems on a causal circuit.

Abramsky and Coecke showed how to formulate quantum theory in a category theoretic framework \cite{AbramskyCoecke} (see also \cite{Coeckeotherstuff}).  This leads to a very rich and beautiful diagramatic theory in which many essential aspects of quantum theory can be understood in terms of simple manipulations of diagrams.  The diagrams can be understood operationally.  Ideas from that work are infused into the present approach.  Indeed, in category theoretic terms, the diagrams in this work can be understood as symmetric monoidal categories.

The ${\bf r}$-${\bf p}$ framework in \cite{Hardy1} is actually simple example of a framework for general probability theories going back originally  to Mackey \cite{Mackey} and has been worked on (and often rediscovered) by many others since including Ludwig \cite{Ludwig}, Davies and Lewis \cite{DaviesLewis}, Araki \cite{Araki}, Gudder \cite{Gudder}, Foulis and Randall \cite{FoulisRandall}.

Barrett elaborated on ${\bf r}$-${\bf p}$ framework in \cite{Barrett}.  He makes two assumptions - that local operations commute and that local tomography is possible (whereby the state of a composite system can be determined by local measurements).  In this work we do not make either assumption.  The first assumption, in any case, would have no content since we are interested in the graphical information in a circuit diagram and interchanging the relative height of operations does not change the graph.  Under these assumptions, Barrett  showed showed that composite systems can be associated with a tensor product structure.  We recover this here for the special case when we have local tomography but the more general case is also studied.   In his paper Barrett shows that some properties which are thought to be specific to quantum theory are actually properties of any non-classical probability theory.

More examples of this nature are discussed in various papers by Barrett, Barnum, Leifer, and Wilce in \cite{BBLW1, BBLW2, BBLW3} and in \cite{BW1, BW2} the general probability framework is further developed.

The assumption of local tomography is equivalent to the assumption that $K_{ab}=K_aK_b$ where $K_{ab}$ is the number of probabilities needed to specify the state of the composite system $ab$ and $K_a$ ($K_b$) is the number needed to describe system $a$ ($b$) alone (this is the content of Theorem 5 below).  Theories having this property were investigated in a paper by Wootters \cite{Wootters} (see also \cite{Wootters1}) in 1990 who showed that they are consistent with the relation $K_a=N_a^r$ where $N_a$ is the number of states that can be distinguished in a single shot measurement (this was used in \cite{Hardy1} as part of the axiomatic structure).

In 1994 Popescu and Rohrlich \cite{PRpaper} exhibited correlations that maximally violate Bell's inequality but do not permit signalling. These correlations are more nonlocal than quantum correlations.   Barrett asked what principles would be required to prescribe such no-signalling correlations to the quantum limit \cite{Barrett}.  Pawlowski {\it et al.} \cite{Pawlowski}  have shown that the Popescu Rohrlich correlations (and, indeed, any correlations more nonlocal than quantum theory) violate a very natural principle they call the information causality principle.  And Gross {\it et al.} \cite{Gross} have shown, as speculated by Barrett \cite{Barrett}, that the dynamics in any theory allowing Popescu Rohrlich correlations are trivial.

Another line of work in this type of framework has been initiated by D'Ariano in \cite{DAriano} who has a set of axioms from which he obtains quantum theory.  In a very recent paper by Chiribella, D'Ariano, and Perinotti \cite{CDP} set up a general probabilistic framework also having the local tomography property.  Like Abramsky, Coecke and co-workers, Chiribella {\it et al.} develop a diagrammatic notation with which calculations can be performed.  They show that theories having the property that every mixed state has a purification have many properties in common with quantum theory.

There have been many attempts at reconstructing quantum theory, not all of them in the probabilistic framework of the sort considered in the above works.  A recent conference on the general problem of reconstructing quantum theory can be seen at \cite{RQTpirsa}.

\section{Essential concepts}

\subsection{Operations and circuits}

\begin{figure*}
\centering
{\includegraphics[width=0.5\textwidth]{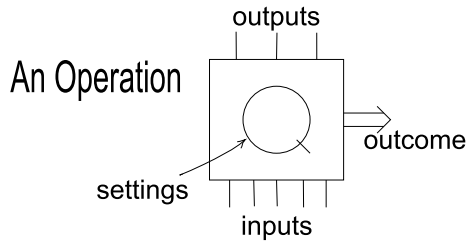}} \caption{\small An operation having knob settings, measurement outcomes, and inputs (at the bottom of box) and outputs (at the top).}\label{operation}
\end{figure*}

The basic building block will be an {\it operation}.  An operation is {\it one use of an apparatus}.  An operation has {\it inputs} and {\it outputs}, and it also has {\it settings} and {\it outcomes} (see Fig.\ \ref{operation}).  The inputs and outputs are apertures which we imagine a system can pass through.  Each input or output can be open or closed. For example, we may close an output by blocking the aperture (we will explain the significance of this later).  The settings may be adjusted by knobs.  The outcomes may be read off a meter or digital display or correspond to a detector clicking or lights flashing for example.  It is possible that there is no outcome readout on the apparatus in which case we can simply say that the set of possible outcomes has only one member.  The same apparatus may be used multiple times in a given experiment.  Each separate use constitutes an operation.

Each input or output is of a given type.  We can think of the type as being associated with the type of system that we imagine passes through.  The type associated with an electron is different than that associated with a photon.  However, from an operational point of view, talk of electrons or photons is a linguistic shortcut for certain operational procedures.  We might better say that the type corresponds to the nature and intended use of the aperture.
Operations can be connected by wires between outputs and inputs of the same type.  These wires do not represent actual wires but rather are a diagrammatic device to show how the apertures on the operations are placed next to one another - this is something an experimentalist would be aware of and so constitutes part of the operational structure.  If we actually have a wire (an optical fibre say) this wire should be thought of as an operation itself and be represented by a box rather than a wire.  Likewise, passage through vacuum also should be thought of as an operation.  The wires show how the experiment is assembled.  Often a piece of self-assembly furniture (from Ikea for example) comes with a diagram showing an exploded view with lines connecting the places on the different parts of the furniture that should be connected. The wires in our diagrams are similar in some respects to the lines in these diagrams (though an experiment is something that changes in time and so the wires represent connections that may be transient whereas the connections in a piece of furniture are static).

There is nothing to stop us trying to match an electron output with a photon input or even a small rock output with an atom input (this would amount to firing small rocks at an aperture intended for individual atoms).  However this would fall outside the intended use of the apparatus and so we would not expect our theory to be applicable (and the apparatuses may even get damaged).

We will often refer to {\it tracing forward} through a circuit.  By tracing forward we mean following a path through the circuit from the output of one operation, along the wire attached, to the input of another operation and then from an output of that operation, along the wire attached, to the input of another operation and so on.  Such paths are analogous to future directed time-like trajectories in spacetime physics.

We require that there can be {\it no closed loops} as we trace forward (i.e. that we cannot get back to the same operation by tracing forward).  This is a natural requirement given that an operation corresponds to a single use of an apparatus (so long as there are no backward in time influences).  It is this requirement of no closed loops that will enable us to foliate.

\begin{figure*}
\centering
{\includegraphics[width=0.8\textwidth]{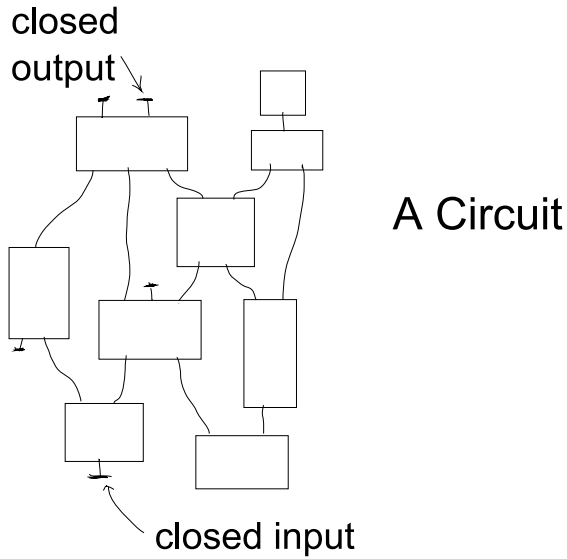}} \caption{\small A bunch of operations wired together form a circuit if there are no open inputs or outputs left over.  We require that there be no closed loops as we trace forward. We have not drawn in the settings or the outcomes (these will usually be taken to be implicit in these circuit diagrams).  There are some closed inputs and outputs.}\label{circuit}
\end{figure*}

In the case that we have a bunch of operations wired together with no open inputs or outputs left over then we will say we have a {\it circuit} (we allow circuits to consist of disconnected parts).  An example of a circuit is shown in Fig.\ \ref{circuit}.

As mentioned above, we assume that any input or output can be closed.  This means that if we have a circuit fragment with open inputs and outputs left over we can simply close them to create a circuit.  This is useful since the mathematical machinery we will set up starts with circuits.  Closing an output can be thought of as simply blocking it off. The usefulness of the notion of closing an output relates to the possibility of having no influences from the future.  This will be discussed in Sec. \ref{closable}.  Closing an input can be thought of as sending in a system corresponding to the type of input in some fiducial state.  We will not make particular use of the notion of closing a input (beyond that it allows us to get circuits from circuit fragments) and so we need not be more specific than this.  We could set up the mathematical machinery in this chapter without assuming that we can close inputs and outputs without much more effort but the present approach has certain pedagogical advantages.

\subsection{Time in a circuit}

We do not think of time as something in the background but rather define it in terms of the circuit itself. We take the attitude that two circuits having the same circuit diagram (in the graphical sense) are equivalent.  Hence there is no physical meaning to sliding operations along wires with respect to some background time.  This is a natural attitude given the interpretation of wires as showing how apertures are placed next to each other rather than as actual wires.

\begin{figure*}
\centering
{\includegraphics[width=0.7\textwidth]{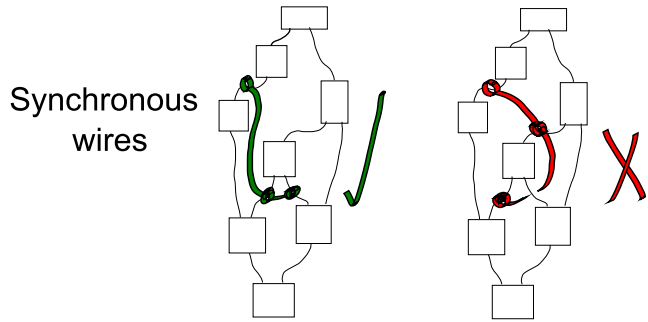}} \caption{\small A set of wires is synchronous if it is not possible to get from one to another by tracing forward.  On the left we see an example of a set of wires which are synchronous and on the right an example of a set which is not.}\label{synchronouswires}
\end{figure*}

We define {\it a synchronous set of wires} to be any set of wires for which there does not exist a path from one wire to another in the set if we trace forward along wires from output to input.  See Fig.\ \ref{synchronouswires} for examples.

We call a synchronous set of wires a {\it hypersurface}, $H$, if it partitions the circuit into two parts, $\gamma_H^-$ and $\gamma_H^+$ that are not connected other than by wires in the hypersurface. Each of the wires in the hypersurface has an end connected to an output (the \lq\lq past") and an end connected to an input (the \lq\lq future").  $\gamma_H^-$ is the part of the circuit to the \lq\lq past" and $\gamma_H^+$ is the part of the circuit to the \lq\lq future" of the hypersurface.  A hypersurface, as defined here, is the circuit analogue of a spacelike hypersurface in spacetime physics.

We say two hypersurfaces are distinct if at least some of the wires are different.
We say that hypersurface $H_2$ is after hypersurface $H_1$ if the intersection of the past of $H_1$ (this is $\gamma_{H_1}^-$) and the future of $H_2$  (this is $\gamma_{H_2}^+$) has no operations in common.  If we can get from every wire in $H_2$ by tracing forward from a wire in $H_1$ then $H_2$ is {\it after} $H_1$ (there are, however, examples of $H_2$ after $H_1$ that are not like this).

A {\it foliation} is a ordered set of hypersurfaces $\{H_t\}$ such that $H_{t+1}$ is after $H_t$. A {\it complete foliation} is a foliation that includes every wire in the circuit.  It is easy to prove that complete foliations exist for every circuit. Define an {\it initial wire} to be one connected to an output of an operation having no open inputs.
Take the set of all initial wires (this cannot be the null set as long as we have at least one connecting wire in this circuit).  These wires form a hypersurface $H_1$.   Consider the set of operations for which these wires form inputs.   Since, according to the wiring constraints, there can be no closed loops, there must exist at least one operation in this set which has no inputs from wires connected to outputs of other operations in this set.  Substitute the wires coming out of one such operation for the wires going into the operation in $H_1$ to form a new spacelike hypersurface $H_2$ (this is after $H_1$).   This can be repeated until all wires have been included forming a complete foliation.  This proves that complete foliations always exist.  There can, of course, exist other complete foliations that are not obtainable by this technique.

Although we do not use a notion of a background time to time-order our operations, it is the case that these foliable structures are consistent with a Newtonian notion of an absolute background time.  Simply choose one complete foliation and take that as corresponding to our Newtonian time.  They are, however, more naturally consistent with relativistic ideas since, for a general circuit, there exist multiple foliations.

\subsection{Probability}\label{closable}

Now we are in a position to introduce probability into the picture.  Probability is a deeply problematic notion from a philosophical point of view \cite{Gillies}.  There are various competing interpretations.  All these interpretations attempt to account for the empirical fact that, in the long run, relative frequencies are stable - that if you toss a coin a million times and get 40\% heads, then if you toss the same coin a million times again, you will get 40\% heads again (more or less).  It is not the purpose of this chapter to solve the interpretational problems of probability and so we will adopt the point of view that probability is a limiting relative frequency.  This gives us the basic mathematical properties of probability:
\begin{enumerate}
\item Probabilities are non-negative
\item Probabilities sum to 1 over a complete set of mutually exclusive events.
\item Bayes rule, $\text{Prob}(A \& B) = \text{Prob}(A|B) \text{Prob}(B)$, applies.
\end{enumerate}
We could equally adopt any other interpretation of probability that gives us these mathematical properties and set up the same theoretical framework.

\begin{figure*}
\centering
{\includegraphics[width=0.6\textwidth]{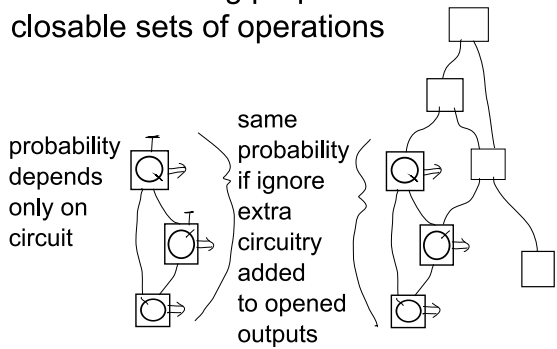}} \caption{\small A set of operations is closable if any circuit built from it has a probability associated with it depending only on that circuit and if that probability is independent of any extra circuitry.}\label{closablef}
\end{figure*}

Typically an experimentalist will have available to him some set of operations, $\cal O$, he can use to build circuits. On each operation in the circuit are various possible settings (among which the experimentalist can choose) and various outcomes one of which will happen.  We say the circuit is setting specified if each operation is given.  We say the circuit is setting-outcome specified if the setting and outcome on each operation is specified.  A setting-outcome specified circuit corresponds to what happens in single run of the experiment.  We define:
\begin{description}
\item[{\bf Closable sets of operations.}]A set of operations, $\cal O$, is said to be closable if, for every setting-outcome specified circuit that can be built from $\cal O$,
\item[~(i)] \ there is a probability depending only on the particulars of this circuit that is independent of choices made elsewhere, and
\item[~(ii)] \!  if we open closed outputs on this circuit and add on extra circuitry then the probability associated with the original bit of setting-outcome specified circuitry (ignoring outcomes associated with the extra bit) is unchanged for any such extra circuitry we can add (see Fig.\ \ref{closablef}).
\end{description}
Part (i) of this definition concerns choices made elsewhere.  These could be choices of settings on operations in other circuits (disjoint from this one), choices of what circuits to build elsewhere, or choices not even having to do with circuits built from the given set of operations.    We might also have said that the probability associated with a setting-outcome specified circuit is independent of the outcomes seen in other circuits.  This is a very natural assumption since otherwise the probability attached to a particular circuit could be different if we restrict our attention to the case where we had seen some particular outcomes on other circuits.  In the case where (i) and (ii) hold and also the probability is independent of the outcomes in other circuits we will say the set of operations is {\it fully closable}.  It turns out we can go a very long way without assuming this.  Further, we will prove that in Section \ref{disjointfrags} that if a very natural condition holds then closable sets are, in any case, fully closable.

Part (ii) imposes a kind of closure from the future - choices on operations only connected to a part of a circuit by outgoing wires (or even choices of what circuitry to place after outgoing wires) do not effect the probabilities for this circuit part.  This could almost be regarded as a definition of what we mean by wires going from output to input.  We do not regard it as a definition though since it corresponds to a rather global property of circuits rather than a property specific to a given wire.

It is interesting to consider examples of sets of operations which are not closable.  Imagine for example that, among his operations, an experimentalist has an apparatus which he specifies as implementing an operation on two qubits but actually it implements an operation on three qubits - there is an extra input aperture he is unaware of.  If he builds a circuit using this gate then an adversary can send a qubit into the extra input which will effect the probability for the circuit.  Thus, the probability would depend on a choice made elsewhere.  In this case we could fix the situation by properly specifying the operation to include the extra input.  The notion of closability is important since it ensures that the experimentalist has full control of his apparatuses.  It is possible, at least in principle, that a set of operations cannot be closed by discovering extra apertures.   By restricting ourselves to physical theories that admit closability we are considering a subset of all possible theories (though a rather important one).

\subsection{Can no-signalling be an axiom?}

\begin{figure*}[t]
\centering
{\includegraphics[width=0.9\textwidth]{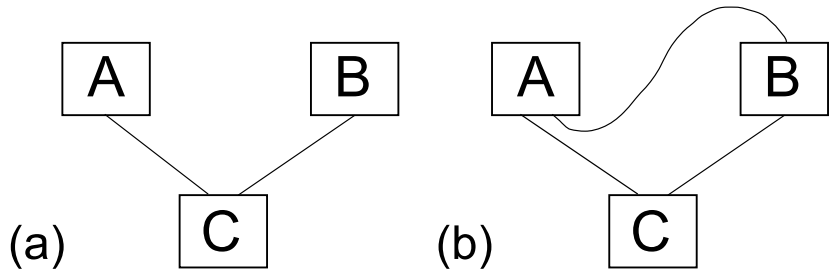}} \caption{\small If there is no-signalling between $A$ and $B$ then the correct circuit is that shown in (a).  However, if there is signalling from $B$ to $A$ then (a) could not be the correct circuit.  Instead, the correct circuit would have to be something like that shown in (b).  The framework is perfectly capable of incorporating signalling.  Hence the assumption of no-signalling is not an assumption of the framework but rather corresponds to asserting that the correct circuit for the given no-signalling situation is the one in (a).}\label{nosignalling}
\end{figure*}

Many authors have promoted no-signalling as an axiom for Quantum Theory. It may appear that part (ii) of the definition of closability sneaks in a no-signalling assumption here.  In fact this is not the case.  Indeed, a no-signalling axiom would actually have limited content in this circuit framework.  Consider the circuit shown in Fig.\ \ref{nosignalling}(a).  A no-signalling axiom would assert that a choice at one end, $B$ say, cannot effect the probability of outcomes at the other end, $A$.  However, this is actually implied by part (ii) since $B$ is connected to the $AC$ part of the circuit by an out-going wire. Hence, it looks like we are assuming no-signalling.  However, this is not an assumption for the framework but only a consequence of asserting that the correct circuit is the one in Fig.\ \ref{nosignalling}(a).  Imagine that there actually was signalling such that probabilities for the $AC$ part of the circuit depended on a choice at $B$ then, under the assumption that the operations are drawn from a closable set, it is clear that the situation cannot be described by Fig.\ \ref{nosignalling}(a).  Rather, the situation would have to be that shown in Fig.\ \ref{nosignalling}(b) where there is an extra wire going from $B$ to $A$ (or something like this but with more structure).  This framework is perfectly capable of accommodating both signalling and no-signalling situations by appropriate circuits and, consequently, we are not sneaking in a no-signalling assumption.

This reasoning leads us to question whether a no-signalling axiom could do any work at all.  Indeed it is often unappreciated in such axiomatic discussions that the usual framework of Quantum Theory does allow signalling.  One can write down nonlocal Hamiltonians which will, for example, entangle product states.  Of course, in Quantum Field Theory one incorporates a no-signalling property by demanding that field operators for space-like separated regions commute so that such nonlocal Hamiltonians are ruled out.  However, this is an example where we have a given background. In general, a no-signalling axiom with respect to some particular given background would restrict the type of circuits we allow. For example the circuit in Fig.\ \ref{nosignalling}(b) would not be allowed unless $A$ was in the future light cone of $B$.  In quantum field theory we have an example where a no-signalling axiom with respect to a Minkowski background restricts the types of unitary evolution and measurement that are allowed.  However, it is often claimed that the abstract Hilbert space framework itself (which makes no mention of Minkowski spacetime) can be derived using no-signalling as one of the axioms.  It is this more ambitious claim we question.  In fact we will see that we can define this abstract framework of quantum theory for any circuit (as long as there are no closed loops) including no-signalling and signalling situations (as in Fig.\ \ref{nosignalling}(a) and (b)).  Hence a no-signalling axiom clearly cannot be regarded as a constraint on this abstract framework.

This criticism of the usefulness of no-signalling as an axiom does not apply to a recently proposed generalisation of this principle in an extraordinary paper by Pawlowski {\it et al.} \cite{Pawlowski}.  They introduced the {\it information causality principle}.  It was shown that this very compelling principle limits violations of the Bell inequality to the quantum limit.  Imagine that Alice receives $n$ classical bits of information.  She communicates $m$ classical bits to Bob.  Bob is expected to reveal the value of one of the $n$ classical bits though neither Alice or Bob know which one this will be in advance.  The information causality principle is that Alice and Bob can only be successful when $n$ is less than or equal to $m$.  For $m=0$ this is the no-signalling assumption we have criticized.  The information causality principle can be read as implying that if the task cannot be achieved for $m=0$ then it cannot be achieved for any other value of $m$.  This principle would be useful in prescribing what is possible in the framework described in this chapter.  Consider two fragments of a circuit that cannot be connected by tracing forward (these fragments are analogous to spacelike separated regions). The information causality principle implies there is no way of accomplishing the above task for any $m$ with $n>m$ between these two circuit fragments.  That we cannot do this for $m=0$ is already implied (assuming that we have the correct circuit).

Although a simple no-signalling \lq\lq across space" principle is of limited use for the above reasons, we do employ what might be regarded as a no-signalling backward in time principle since we do not allow closed loops in a circuit.


\subsection{Systems and composite systems}

We wish to give an operational definition of what we mean by the notion of a {\it system}.  We may find that whenever we press a button on one box, a light goes on on another box.  We can interpret this in terms of a system passing between the two boxes.  We find this happens only when we place the boxes in a certain arrangement with respect to one another (which we think of as aligning apertures).  Given this we clearly want to associate systems with wires.  Hence, we adopt the following definition:
\begin{quote}
{\bf A system} of type $abc\ldots$ is, by definition, associated with any set of synchronous wires of type $a, b, c, \ldots$ in any circuit formed from a closable set of operations.
\end{quote}
We may refer to a system type corresponding to more than one wire by a single letter.  Thus we may denote the system type $abc$ by $d$.

The usefulness of the notion of closable sets of operations is it that it leads to  wires being associated with the sort of correlation we expect for systems given our usual intuitions about what systems are.  Nevertheless, our definition of system is entirely operational since wires are defined operationally.

It is common to speak of composite systems.  We define a composite system as follows:
\begin{quote}
{\bf A composite system}, $AB$, is associated with any two systems (each associated with disjoint sets of wires) if the union of the sets of wires associated with system $A$ and system $B$ forms a synchronous set.
\end{quote}
This definition generalizes in the obvious way for more than two systems.  A system of type $aabc$ can be regarded as a composite of systems of type $aa$ and $bc$ or a composite of systems of type $aac$ and $b$, or a composite of systems of type $ac$, $a$ and $b$ to list just a few possibilities.  Systems associated with a single wire cannot be regarded as composite.

A hypersurface consists of synchronous wires and so can be associated with a system (or composite system).  A complete foliation can therefore be associated with the evolution of a system through the circuit (though the system type can change after each step).  This evolution can also be viewed as the evolution of a composite system.

\section{Preparations, transformations, and effects}

\subsection{Circuit fragments}

\begin{figure*}
\centering
{\includegraphics[width=0.95\textwidth]{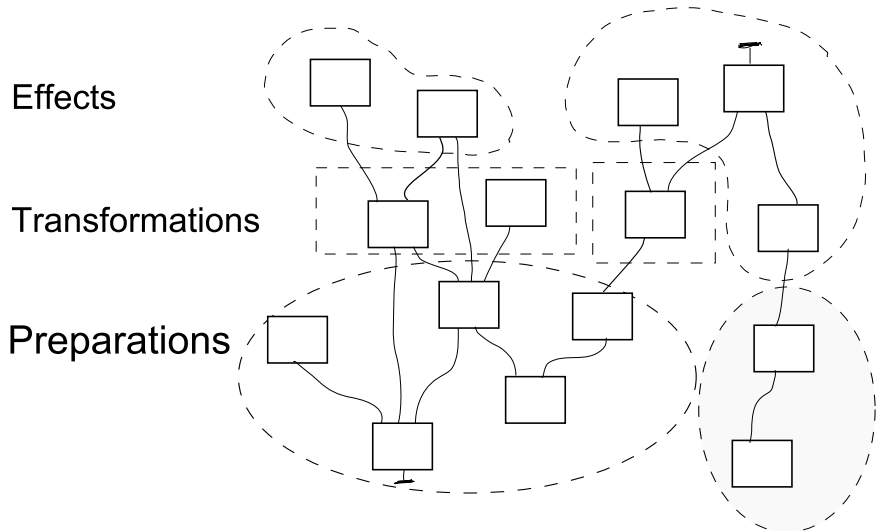}} \caption{\small A circuit fragment is a part of the circuit having inputs  wires in a synchronous set and output wires also in a synchronous set.  A circuit can be divided up many ways into circuit fragments corresponding to preparations (no open inputs), transformations, and effects (no open outputs).}\label{preptranseff}
\end{figure*}

We can divide up a circuit into fragments corresponding to preparations, transformations, and effects as shown in Fig.\ \ref{preptranseff}.  By the term {\it circuit fragment} we mean a part of a circuit (a subset of the operations in the circuit along with the wires connecting them) having inputs coming from a synchronous set of wires and outputs going into a synchronous set of wires.  We allow lone wires in a circuit fragment (wires not connected to any operations in the fragment).  An example of a lone wire is see in the circuit fragment in the rectangle on the left in Fig.\ \ref{preptranseff}.  The lone wire corresponds to the identity transformation on that system and contributes an input and output to the circuit fragment.  Generally, we take the term {\it circuit fragment} to imply that the settings and outcomes at each operation associated with these circuit fragments have been specified. A circuit fragment is, essentially, an operation at a course grained level.  {\it Preparations} correspond to a circuit fragment having outputs but no open inputs.   {\it Transformations} have inputs and outputs.  {\it Effects} have inputs but no open outputs .  Note that preparations and effects are special cases of transformations.

\subsection{States}

A preparation prepares a system.  For a given type of system there will be many possible preparations.  We will label them with $\alpha$. This label tells us what circuit fragment is being used to accomplish the preparation (including the specification of the knob settings and outcomes on each operation).  Associated with each preparation for a system of type $a$ will be a state (labeled by $\alpha\in \text{Prep}_a$).  We can build a circuit having this preparation by adding an effect for a system of type $a$.  There are many possible effects labeled by $\beta\in \text{Eff}_a$. The label $\beta$ tells us what circuit fragment is used to accomplish the effect along with the knob settings and outcomes at each operation.  Associated with the circuit is a probability $p^{\alpha\beta}$. We define the {\it state} associated with preparation $\alpha$ to be that thing represented by any mathematical object that can be used to calculate $p^{\alpha\beta}$ for all $\beta$.  We could represent the state by the long list
\begin{equation}
{\bf P}^\alpha_a = \left( \begin{array}{c} \vdots \\ p^{\alpha\beta} \\ \vdots \end{array} \right) ~~~ \beta\in \text{Eff}_a
\end{equation}
However, in general, physical theories relate physical quantities.  Hence, it is only necessary to list a subset of these quantities where the remaining quantities can be calculated by the equations of the physical theory.  We call the forming of this subset of quantities {\it physical compression}.  In the current case, we expect the probabilities in this list to be related.  We consider the maximum amount of physical compression that is possible by linear means.  Thus we write the state as
\begin{equation}
{\bf p}^\alpha_a = \left( \begin{array}{c} \vdots \\ p^{\alpha\beta} \\ \vdots \end{array} \right) ~~~ \beta\in \Omega_a\subseteq \text{Eff}_a
\end{equation}
where there exist vectors ${\bf r}^\alpha_a$ such that
\begin{equation}
p^{\alpha\beta}= {\bf r}^\beta_a \cdot {\bf p}^\alpha_a ~~\text{for all}~ \alpha\in \text{Prep}_a, ~\beta\in\text{Eff}_b
\end{equation}
We call the set $\Omega_a$ the fiducial set of effects for a system of type $a$. The choice of fiducial set need not be unique - we simply make one choice and stick with it.  Since we have applied as much linear physical compression as possible $|\Omega_a|$ is the minimum number of probabilities required to calculate all the other probabilities by linear equations.  The vectors ${\bf r}^\beta_a$ are associated with effects on a system of type $a$.  For the fiducial effects they consist of a 1 in the $\beta$ position and 0's elsewhere.

An important subtlety here is that we define states in terms of joint rather than conditional probabilities.  This makes more sense for the circuit model since, generally, we want to calculate a probability for a circuit.  If we want to calculate conditional probabilities we can use Bayes' rule in the standard way.

\subsection{Transformations}

Now consider a preparation $\alpha$ which prepares a system of type $a$ in state ${\bf p}^\alpha_a$ followed by a transformation $\beta$ which outputs a system of type $b$.  We can regard the preparation and transformation, taken together, as a new preparation $\alpha\beta$ for a system of type $b$ with state ${\bf p}^{\alpha\beta}_b$.  Now follow this by a fiducial effect $\gamma\in\Omega_b$. The probability for the circuit $\alpha\beta\gamma$ can be written
\begin{equation}
p^{\alpha\beta\gamma} = {\bf r}^\gamma_b \cdot {\bf p}^{\alpha\beta}_b = {\bf r}^{\beta\gamma}_a\cdot{\bf p}^\alpha_a
\end{equation}
where ${\bf r}^{\beta\gamma}_a$ is the effect vector associated with the measurement consisting of the transformation $\beta$ followed by the effect $\gamma$.  Given the special form of the fiducial effect vectors it follows that the state transforms as
\begin{equation}
{\bf p}^{\alpha\beta}_b =  {}^b\!Z^{\beta}_a {\bf p}^\alpha_a
\end{equation}
where $^b\!Z^\beta_a$ is a $|\Omega_b|\times |\Omega_a|$ matrix such that its $\beta$ row is given by the components of the effect vectors ${\bf r}^{\beta\gamma}$.  We use a subscript, $a$, for the inputted system type and a pre-superscript, $b$, for the outputted system type. Hence, a general transformation is given by a matrix acting on the state.  If the matrix transforms from one type of system to another type of system with a different number of fiducial effects then it will be rectangular.

The general equation for calculating probabilities is
\begin{equation}
p_{\alpha\beta\gamma}= ({\bf r}^\gamma_b)^T {}^b\!Z^\beta_a {\bf p}^\alpha_a
\end{equation}
where $T$ denotes transpose.  If we have more than one transformation then, by a clear extrapolation of the above reasoning, we can write
\begin{equation}
p^{\alpha\beta\gamma\delta}= ({\bf r}^\delta_c)^T {}^c\!Z^\gamma_b {}^b\!Z^\beta_a {\bf p}^\alpha_a
\end{equation}
and so on.  Now since the $Z$ matrices can be rectangular we can think of ${\bf p}^\alpha_a$ and $({\bf r}^\gamma_a)^T$ as instances of a transformation matrix. The state ${\bf p}^\alpha_a$ can be thought of as corresponding to the transformation which turns a null system (no system at all) into a system outputted by this preparation and we can change our notation to ${}^a\!Z^\alpha$ instead (a column vector being a special case of a rectangular matrix).  Likewise we can change our notation for the row vector $({\bf r}^\delta_a)^T$ to $Z^\delta_a$ (a row vector being a special case of a rectangular matrix). Then we can write
\begin{equation}\label{ZZZexpression}
p^{\alpha\beta\gamma\delta}= Z^\delta_c \, {}^c\!Z^\gamma_b\, {}^b\!Z^\beta_a \, {}^a\!Z^\alpha
\end{equation}
The agreement of output and input system types is clear (by matching pre-superscript with subscripts between the $Z$'s).

The label $\alpha$ labels the circuit fragment along with the knob settings and the outcomes. Sometimes it is useful to break these up into separate labels.  Thus we write
\begin{equation}
\alpha \equiv ({\mathcal F}, \varphi, l)
\end{equation}
where ${\mathcal F}$ denotes the circuit fragment before the settings and outcomes are specified, $\varphi$ denotes the settings on the operations, and $l$ denotes the outcomes.  In particular, this means that we can notate the effects associated with the different outcomes of a measurement as $\alpha_i = ({\mathcal F}, \varphi, l_i)$.

Let $L$ be the set of possible outcomes $l_i$ for a given measurement (with fixed $\cal F$ and $\varphi$).  We can subdivide the set $L$ into disjoint sets $L_k$ where $\cup_k L_k = L$.   We could choose to be ignorant of the actual outcome $l_i$ and rather only record which set $L_k$ it belongs to.  In this case the transformation effected can be denoted $\overline{\alpha}_k$.  Since we have used linear compression, we must have
\begin{equation}
{}^b\!Z_a^{\overline{\alpha}_k} = \sum_{ i\in I_k} {}^b\!Z_a^{\alpha_i}
\end{equation}
where $I_i$ is the set of $i$'s corresponding to the $l_i$'s in $L_k$.   Since we can always choose to be ignorant in this way, we must include such transformations in the set of allowed transformations.

The matrices corresponding to the set of allowed transformations must be such that, when closed expressions such as
(\ref{ZZZexpression}) are calculated, they always give probabilities between 0 and 1.  This is an important constraint on this framework.

\subsection{The identity transformation}

One transformation we can consider is where we do nothing.  The wires coming in are the same as the wires coming out and no operation has intervened.  We will denote this transformation by $0$.  Then we have, for example,
\begin{equation}
{}^{a}\!Z_{a}^{0}
\end{equation}
This is a $|\Omega_{a}|\times |\Omega_{a}|$ matrix and must be equal to the identity since, as long as it is type matched, it can be inserted as many times as we like into any expression where non-trivial transformations act also.

\subsection{The trace measurement}

One effect we can perform on a preparation $\alpha$ is to close all outputs.  This forms a circuit and hence there is an associated probability, $p^{\alpha -}$ (where $-$ denotes that the outputs have been closed). This is an effect and hence we must have
\begin{equation}
p^{\alpha -}={\bf r}^-_a \cdot {\bf p}_a^\alpha
\end{equation}
where the vector ${\bf r}^-_a$ corresponds to this effect.  We call this the trace measurement (terminology borrowed from quantum theory where this effect corresponds to taking the trace of the density matrix).  It follows from part (ii) of the condition for a closable set of operations that this is the probability associated with the preparation part of the circuit even if the outputs are open and more circuitry is added.

In the case that ${\bf r}^-_a \cdot {\bf p}_a^\alpha =1$ we say that the state is of {\it norm one}.  In general we do not expect states to be of norm one since they consist of joint rather than conditional probabilities and hence we require only that $0 \leq {\bf r}^-_a \cdot {\bf p}_a^\alpha \leq 1$.

We can {\it normalize} a state by dividing by ${\bf r}^-_a \cdot {\bf p}_a^\alpha$.  We denote the normalized state by
\begin{equation}
\bar{\bf p}_a^\alpha \equiv \frac{{\bf p}_a^\alpha}{{\bf r}^-_a \cdot {\bf p}_a^\alpha}
\end{equation}
We cannot guarantee that $\bar{\bf p}_a^\alpha$ belongs to the set of allowed states (i.e. that there exists a state for which ${\bf p}_a^\alpha=\bar{\bf p}_a^\alpha$) since preparations for all states parallel to $\bar{\bf p}_a^\alpha$ may be intrinsically probabilistic.

\section{Mixtures}

\subsection{Forming Mixtures}

Imagine that we have a box with a light on it that can flash and an aperture out of which a system of type $a$ can emerge.  With probability $\lambda_i$ we place preparation $\alpha_i$ in the box such that the system (which we take to be of type $a$) will emerge out of the aperture and such that the light will flash if the outcomes corresponding to this preparation are seen.  The state prepared for this one $i$ is $\lambda_i{\bf p}^{\alpha_i}_a$. If $\lambda_i =0$ then the state prepared is the {\it null state} ${\bf 0}_a$ which has all 0's as entries.  If we use this box for a set $\alpha_i$  with $i\in I$ such that $\sum_{i\in I}\lambda_i \leq 1$ then the state prepared is
\begin{equation}
\sum_{i\in I} \lambda_i {\bf p}_a^{\alpha_i}
\end{equation}
This is a linear sum of terms since we have linear compression. This process of using a box may be beyond the experimental capacities of a given experimentalist.  It certainly takes us outside the circuit model as previously described.  However, we can always consider taking mixtures like this at a mathematical level.

A technique that can be described in the circuit model is the following.  Consider placing a single preparation circuit into the box described above where $\alpha_j = ({\mathcal F}, \varphi, l_j)$ where $l_j$ labels the outcomes.  We can arrange things so that the light flashes only if $j\in J$ (where $J$ is some subset of the $j$'s).  The state is then given by
\begin{equation}
\sum_{j\in J} {\bf p}_a^{\alpha_j}
\end{equation}
This technique does not require having a coin to generate probabilities $\lambda_i$ and neither does it require the placing of different circuit fragments into a box depending on the outcome of the coin toss.

The most general thing we can do is a mixture of the two above techniques. With probability $\lambda_i$ we place a circuit ${\mathcal F}_i$ with settings $\varphi_i$ and outcomes $l_{ij}$ in the box for $i\in I$ and $j\in J_i$.  The state we obtain is
\begin{equation}
\sum_{i\in I, j\in J_i} \lambda_{i} {\bf p}^{\alpha_{ij}}_a
\end{equation}
We can absorb the $\lambda$'s into the ${\bf p}$'s and relabel so we obtain that a general mixture is given by
\begin{equation}
{\bf p}^\alpha = \sum_i {\bf p}^{\alpha_i}_a
\end{equation}
We have the constraints that ${\bf r}_a^-\cdot {\bf p}_a^{\alpha_i} \geq 0$ and $\sum_i {\bf r}_a^-\cdot {\bf p}_a^{\alpha_i} \leq 1$.

Note that if we write ${\bf p}_a^{\alpha_i}\equiv \mu_i \bar{\bf p}^{\alpha_i}_a$ where $\bar{\bf p}^{\alpha_i}_a$ is normalized, and include an extra state ${\bf p}^0_a\equiv {\bf 0}$ (the null state) then we have
\begin{equation}
{\bf p}^\alpha = \sum_i \mu_i\bar{\bf p}^{\alpha_i}_a ~~~\text{where}~~~ \mu_i \geq 0, ~~~\sum_i \mu_i = 1
\end{equation}
where the sum now includes the null state.   Hence we can interpret a general mixture as a convex combination of normalized states and the null state.

\subsection{Homogeneous, pure, mixed, and extremal states}

If two states are parallel then they give rise to the same statistics up to an overall weighting and if we condition on the preparation then they have exactly the same statistics.  We define a {\it homogeneous state} as one which can only be written as a sum of parallel states.  Thus ${\bf p}_a^\alpha$ is homogeneous if, for any sum,
\begin{equation}
{\bf p}_a^\alpha = \sum_{i\in I} {\bf p}_a^{\alpha_i}
\end{equation}
we have ${\bf p}_a^{\alpha_i} =\eta_{ii'} {\bf p}_a^{\alpha_{i'}}$ for all $i,i'\in I$.  A state which is not homogeneous (i.e. which can be written as a sum of at least two non-parallel states) is called a {\it heterogeneous state}.

Given a particular homogeneous state, there will, in general, be many others which are parallel to it but of different lengths.  We call the longest among these a {\it pure state}.

A {\it mixed state} is defined to be any state which can be simulated by a probabilistic mixture of distinct states in the form $\sum_j \lambda_j {\bf p}_a^{\alpha_j}$ where $\lambda_j\geq 0$ and $\sum_j\lambda_j=1$.  A pure state is not a mixture (since the ${\bf p}_a^{\alpha_j}$'s would have to parallel to the given pure state, and therefore, given the $\sum_j \lambda_j=1$ condition, equal to the given pure state).   Homogeneous states which are not pure are mixtures.  Heterogeneous states are also mixtures.  {\it Extremal states} are defined to be states which are not mixtures.  Pure states are extremal.  The null state is also extremal.  If all pure states have norm equal to one (i.e. ${\bf r}_a^-\cdot {\bf p}_a^\alpha= 1$) then there are no more extremal states beyond the pure states and null state.  However, if some pure states have ${\bf r}_a^-\cdot {\bf p}_a^\alpha< 1$, then there may be additional extremal states.

Usually treatments of convex sets of states do not make these distinctions.  More care is necessitated here because states are based on joint rather than conditional probabilities.

Any state, extremal or mixed, can be written as the sum of homogeneous states.  This means that there must exist at least one set of $|\Omega_a|$ linearly independent homogenous states all of which can be pure. There cannot exist sets with more linearly independent states than this.


\subsection{Optimality of linear compression}

We state the following theorem
\begin{quote}
{\bf Theorem 1}~ If we allow arbitrary mixtures of preparations then (1) linear compression is optimal and (2) optimal compression is necessarily linear.
\end{quote}
The first point follows since there must exist $|\Omega_a|$ linearly independent states (otherwise we could have implemented further linear compression).  We can take an arbitrary mixture with $\sum_i \lambda_i \leq 1$ then these $\lambda$'s are all independent and hence we need $|\Omega_a|$ parameters and the compression is optimal. To prove the second point consider representing the state by a list of $|\Omega_a| $ probabilities $p_a^{\alpha\beta'}$ with $\beta'\in \Omega'_a$ where we do not demand that a general probability is given by a linear function of these probabilities.  Represent this list by the vector ${\bf q}^\alpha_a$.  Now
\begin{equation}
p^{\alpha\beta'} = {\bf r}_a^{\beta'} \cdot {\bf p}_a^\alpha ~~~ \text{for all} \beta'\in \Omega'_a
\end{equation}
Hence ${\bf q}^\alpha_a = C {\bf p}_a^\alpha$ where $C$ is a square matrix with real entries.   $C$ must be invertible since otherwise we could specify ${\bf q}_a^\alpha$ with fewer than $|\Omega_a|$ probabilities.  Hence
\begin{equation}
p^{\alpha\beta} = {\bf r}^\beta_a \cdot C^{-1} {\bf p}_a^\alpha
\end{equation}
for a general $\beta$. Hence the probability is linear in ${\bf q}_a^\alpha$ and so the compression is linear.



\section{Composition}

\subsection{Preliminaries and notation}

As we discussed above, systems associated with more than one wire can be thought of as composite.  The ${\bf p}$, $\bf r$, $Z$ framework just discussed can be enriched by adapting it to deal separately with the components of composite systems (rather than regarding all the wires at each time step as constituting a single system). The advantage of this is that we can break up the calculation into smaller parts and thereby define a theory by associating matrices to smaller transformations.  Ultimately, we would like to have a matrix associated with each operation in the set of allowed operations which can be used to calculate the probability for any circuit.  A transformation is now associated with a matrix such as
\begin{equation}
{}^{bacd}\!Z_{acb}^{\alpha}
\end{equation}
This transformation inputs a system of type $acb$ and outputs a system of type $bacd$. The label $\alpha$ denotes the circuit fragment used to do this including the knob settings and the outputs at each operation in the fragment. The ordering of the symbols representing the system (such as $bacd$) is significant in that it is preserved  between transformations to indicate how the wires are connected.  Thus, the matrix for a transformation comprised of two successive transformations is  written
\begin{equation}\label{twotransformations}
{}^d\!Z_{bacd}^\beta \, {}^{bacd}\!Z_{acb}^{\alpha}
\end{equation}
The system types in between the two transformations must match (as in this example) since we employ the convention that the wires are in the same order from the output of one transformation to the input of the next.  We can allow that the symbols for the systems (such as $c$) actually correspond to composite systems (so that they correspond to a cluster of wires).  For example, it might be that $c=aabb$.
Some transformations consist of disjoint circuit fragments and it is useful to have notation for these.  We write
\begin{equation}
{}^{(b)(ac)(-)}\!Z_{(a)(c)(b)}^{(\alpha)(\beta)(\gamma)}
\end{equation}
to indicate that the transformation consists of three disjoint parts, one transforming from system type $a$ to $b$ and labeled by $\alpha$, one transforming from $c$ to $ac$ and labeled by $\beta$ and which inputs $b$ and has no open outputs (which we denote by $-$ when necessary for disambiguation) and labeled by $\gamma$.  If it is clear from the context we will sometimes use the less cumbersome notation
\begin{equation}
{}^{cd}\!Z_{ab}^{\alpha\beta} \equiv {}^{(c)(d)}\!Z_{(a)(b)}^{(\alpha)(\beta)}
\end{equation}
We may sometimes want to depart from the convention that the wires are in the same order from one transformation to the next in which case we label the wires (and wire clusters as appropriate) using intergers $1, 2, \dots$ as follows
\begin{equation}
{}^{(b)_5(ac)_{64}(d)_7}\!Z_{(a)_1(c)_3(b)_2}^{(\alpha)(\beta)(\gamma)}
\end{equation}
In this example wire 6 is an output wire of type $a$.  We can then rewrite (\ref{twotransformations}) as
\begin{equation}
{}^{(d)_8}\!Z_{(dcab)_{7654}}^{(\beta)} \, {}^{(bacd)_{4567}}\!Z_{(acb)_{123}}^{(\alpha)}
\end{equation}
We see that the wires match (for example wire 6 is of type $c$ as an output from the first transformation and an input into the second transformation).  The integers labeling the wires are of no significance and any expression is invariant under any reassignment of these labels (this is a kind of discrete general covariance).

\subsection{Commutation}\label{commutationsection}

\begin{figure*}[t]
\centering
{\includegraphics[width=0.5\textwidth]{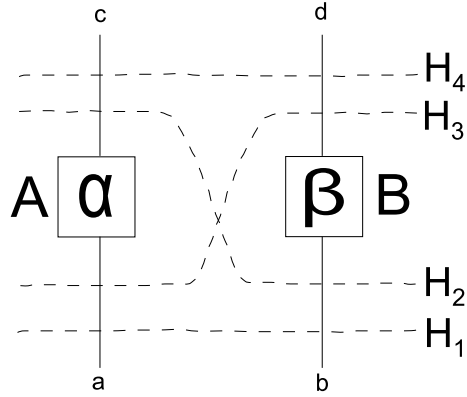}} \caption{\small We can consider the evolution of this circuit with respect to different foliations.}\label{commutingboxes}
\end{figure*}

Consider the situation shown in Fig.\ \ref{commutingboxes}.  By inspection of this diagram we can write
\begin{equation}\label{gencommutation}
{}^{cd}\!Z_{ab}^{\alpha\beta} = {}^{cd}\!Z_{cb}^{0\beta} \, {}^{cb}\!Z_{ab}^{\alpha 0}
= {}^{cd}\!Z_{ad}^{\alpha 0} \, {}^{ad}\!Z_{ab}^{0\beta}
\end{equation}
where the $0$ denotes that we do nothing (the identity transformation).  The first equation here is obtained by first transforming from hyperplane $H_1$ to $H_2$ (past operation $A$) and then from $H_2$ to $H_4$ (past operation $B$). To get the second equation we evolve from $H_1$ to $H_3$ first (past $B$) then from $H_3$ to $H_4$ (past $A$).   There are a few points of interest here.  First note that we can break down a compound transformation into its parts.  Second, we see that there is a kind of commutation - it does not matter whether we update at $A$ or $B$ first. This property is not assumed but derived from more basic assumptions and definitions.  In the special case where wires $c$ and $d$ are of type $a$ and $b$ respectively (so the transformations do not change the system type) we have the commutation property
\begin{equation}
[ {}^{ab}\!Z_{ab}^{0\beta}, {}^{ab}\!Z_{ab}^{\alpha 0}] \equiv
{}^{ab}\!Z_{ab}^{0\beta} \, {}^{ab}\!Z_{ab}^{\alpha 0} - {}^{ab}\!Z_{ab}^{\alpha 0} \, {}^{ab}\!Z_{ab}^{0\beta} = 0
\end{equation}
The usual commutation relation is, then, a special case of the more general relation in (\ref{gencommutation}) where the local transformations may change the system type.

The fact that we can break up a compound transformation into smaller parts is potentially useful.  But there is a stumbling block.  The matrix ${}^{cb}\!Z_{ab}^{\alpha 0}$ transforms past the $A$ operation. However, it is a $|\Omega_{cb}|\times |\Omega_{ab}|$ matrix.  That is, we still have to incorporate some baggage because we include wire $b$.  It would be good if we could write
\begin{equation}\label{withidentity}
{}^{cb}\!Z_{ab}^{\alpha 0} = {}^{c}\!Z_{a}^{\alpha} \otimes {}^{b}\!Z_{b}^{0}  ~~~~~~~ ?
\end{equation}
where ${}^{b}\!Z_{b}^{0}$ is just the identity matrix (as discussed above).  By considering the sizes of the matrices it is clear that (\ref{withidentity}) implies $|\Omega_{ab}|=|\Omega_a||\Omega_b|$.   In Sec.\ \ref{locality} we show that equation (\ref{withidentity}) holds in general if $|\Omega_{ab}|=|\Omega_a||\Omega_b|$ is true (for any system types $a$ and $b$). We will also see that this condition corresponds to a very natural class of physical theories.  If  (\ref{withidentity}) holds true we can break any circuit down into its basic operations appending the identity transformation as necessary.  Then we can calculate the probability associated with any circuit from the transformation matrices associated with the operations.


\subsection{Homogenous states and composite systems}

Consider a composite system $12$ consisting of systems of types $a$ and $b$ with preparation $C$ labeled $\gamma$.  The state prepared by this is ${\bf p}_{ab}^{\gamma}$.  If we block the output $2$ of the preparation then we have a preparation for a system of type $a$.  Let the state so prepared be ${\bf p_a^\gamma}$.  Even if we do not block output $2$ it follows from part (ii) of the condition for closable sets of operations that this state ${\bf p}_a^\gamma$ gives us the correct probabilities for all measurements on system $1$ alone (that do not involve system $2$).  We call ${\bf p}_a^\gamma$ the {\it reduced state} for system $1$.  It is, effectively, the state of system $1$ taken by itself.

\begin{figure*}
\centering
{\includegraphics[width=0.4\textwidth]{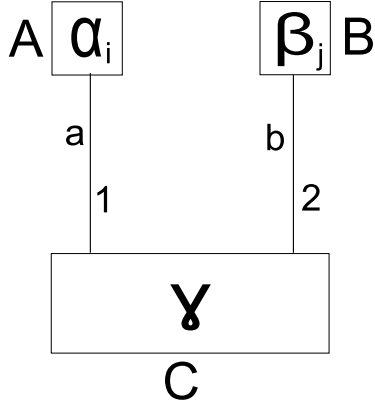}} \caption{\small A bipatite system $12$ of type $ab$ is prepared by some preparation $\gamma$ and subjected to effects $\alpha_i$ and $\beta_i$ on subsystem $1$ and subsystem $2$ respectively.}\label{bipartite}
\end{figure*}

We prove the following theorem.
\begin{quote}
{\bf Theorem 2}~  If one component of a bipartite system is in a homogeneous state (i.e. the reduced state for this system is homogeneous) all joint probabilities for separate effects measured on the two systems factorize.
\end{quote}
Systems $1$ and $2$ can be subjected to measurements $A$ and $B$ respectively (see Fig.\ \ref{bipartite}).  We also consider the possibility of closing either or both outputs from $C$.  Let measurement $A$ have outcomes $l_i$, $i\in I$ the effects for which are labeled $\alpha_i\equiv ({\mathcal F}^A, \varphi^A, l^A_j)$.  Similarly, for $B$ we have outcomes $l^B_j$ for $j\in J$ and effects labeled by $\beta_i\equiv ({\mathcal F}^B, \varphi^B, l^B_j)$.
By part (ii) of the condition for a closable set of operations we have
\begin{equation} \label{sumcc}
p^{\gamma--} = \sum_{i\in I} p^{\gamma \alpha_i -} = \sum_{j\in J} p^{\gamma-\beta_j}
\end{equation}
and
\begin{equation}\label{sumco}
p^{\gamma-\beta_j} = \sum_{i\in I} p^{\gamma\alpha_i\beta_j}
\end{equation}
If the output $2$ from $C$ is closed then we say that the state prepared (for the system of type $a$) is ${\bf p}_a^{\gamma}$ (the reduced state).  We say that the preparation due to $C$ and $B$ with outcome $l_j^B$ (this is a preparation circuit fragment) is ${\bf p}^{\gamma\beta_j}_a$.  It follows from part (ii) of the condition for a closable set of operations that
\begin{equation}
{\bf p}_a^\gamma = \sum_{j\in J} {\bf p}_a^{\gamma\beta_j}
\end{equation}
If ${\bf p}_a^\gamma$ is homogeneous then all the ${\bf p}_a^{\gamma\beta_j}$ states must be parallel to it.  Hence we can write ${\bf p}_a^{\gamma\beta_j}= \eta_j {\bf p}_a^\gamma$.  We must have
\begin{equation}
p^{\gamma\alpha_i\beta_j} = {\bf r}_a^{\alpha_i} \cdot{\bf p}_a^{\gamma\beta_j} = \eta_j {\bf r}_a^{\alpha_i} \cdot {\bf p}_a^{\gamma} = \eta_j p^{\gamma\alpha_i -}
\end{equation}
Summing this over $i$ and using (\ref{sumcc}, \ref{sumco}) we obtain $p^{\gamma-\beta_j}=\eta_j p^{\gamma--}$.  Hence
\begin{equation}\label{gammafactorise}
p^{\gamma\alpha_i\beta_j} p^{\gamma --} = p^{\gamma \alpha_i -} p^{\gamma - \beta_j}
\end{equation}
Here $p^{\gamma --}$ is the probability of the preparation being successful.
Dividing this through by $(p^{\gamma--})^2$ and using Bayes' rule we obtain
\begin{equation}
\text{prob}(l^A_i l^B_j | \text{prep}) = \text{prob}(l^A_i | \text{prep}) \text{prob}(l^B_j | \text{prep})
\end{equation}
Hence we see that if one system is in a homogeneous state then joint probabilities factorize between the two ends (obviously the result also holds if both components are in homogeneous states).  An obvious corollary is
\begin{quote}
{\bf Corollary 1}~
If the state of a bipartite system $12$ of type $ab$ with preparation $\gamma$ is of norm one and the reduced state of either or both components is pure then
\begin{equation}
p^{\gamma\alpha\beta} = p^{\gamma\alpha}p^{\gamma\beta}
\end{equation}
where $\alpha$ is any effect on $1$ alone and $\beta$ is any effect on $2$ alone.
\end{quote}
This is true since pure states are necessarily homogeneous and because $p^{\gamma--}=1$ since the state is of norm 1.

Now we consider a related but slightly different situation. Imagine we have a preparation consisting of two disjoint parts one of which prepares a homogeneous state.
\begin{quote}
{\bf Theorem 3}~
If a preparation, $\gamma\delta$, consists of two disjoint circuit fragments $\gamma$ and $\delta$ which prepare systems of type $a$ and $b$ respectively, and one of these circuit fragments, $\gamma$, taken by itself prepares a homogeneous state, ${\bf p}_a^\gamma$, then the state prepared by closing the outputs of the second circuit fragment of $\gamma\delta$ is parallel to ${\bf p}_a^{\gamma}$ (and therefore also homogeneous).
\end{quote}
The proof of this theorem is based on the same idea as the previous theorem.  We can put $\delta=({\cal{F}}, \varphi, l)$ where $\cal{F}$ denotes the actual circuit fragment, $\varphi$ the settings, and $l$ the outcomes.  Then we can put $\bar{\delta} = ({\cal{F}}, \varphi, \bar{l} )$.  This is the circuit fragment associated with not seeing $l$. Either $l$ or $\bar{l}$ must happen, and hence
\begin{equation}
{\bf p}_a^\gamma = {\bf p}_a^{\gamma\delta} + {\bf p}_a^{\gamma\bar{\delta}}
\end{equation}
Since ${\bf p}_a^\gamma$ is homogeneous both ${\bf p}_a^{\gamma\delta}$ and ${\bf p}_a^{\gamma\bar{\delta}}$ must be parallel to it and hence are also homogeneous.

\subsection{Fiducial measurements for composite systems}\label{fidcomp}

Assume that system $1$ is prepared by some preparation$\gamma$ in a homogeneous state ${\bf p}_a^\gamma$ and, similarly, system $2$ is prepared in homogeneous state ${\bf p}_b^\delta$.  Assume these are two separate preparations corresponding to two disjoint circuit fragments. Hence we can consider the joint preparation $\gamma\delta$ giving rise to the state ${\bf p}^{\gamma\delta}_{ab}$.

A natural question is, what is the relationship between the states ${\bf p}_a^\gamma$, ${\bf p}_b^\delta$, and ${\bf p}^{\gamma\delta}_{ab}$? By virtue of Theorem 3 we know that the reduced state at either end is homogeneous and hence, by virtue of Theorem 2 (equation (\ref{gammafactorise}) in particular) and the fact that the subsystems are in homogeneous states we can write
\begin{equation}\label{fidcompeqn}
{\bf r}_{ab}^{\alpha\beta}\cdot {\bf p}_{ab}^{\gamma\delta} = p^{\gamma\delta\alpha\beta} = \mu_{\gamma\delta} p^{\gamma\alpha} p^{\delta\beta} = \nu_{\gamma_\delta}({\bf r}_a^\alpha \otimes{\bf r}_b^\beta)\cdot ( {\bf p}_a^{\gamma}\otimes {\bf p}_b^{\gamma})
\end{equation}
where we obtain
\begin{equation}\label{strange}
\nu_{\gamma\delta}= \frac{p^{\gamma\delta --}}{p^{\gamma-}p^{\delta-}}
\end{equation}
by putting $\alpha=-$ and $\beta=-$.

We know that there must exist $K_a\equiv |\Omega_a|$ linearly independent homogeneous states for system $1$ (they can all be chosen to be pure).  Let $\gamma\in \overline{\Omega}_a$ be the preparations associated with one such set of linearly independent homogeneous states for system $1$ (where $|\overline{\Omega}_a|= |\Omega_a|$).  Likewise we have $K_b\equiv |\Omega_b|$ linearly independent homogeneous states for system $2$ with preparations $\delta\in \overline{\Omega}_b$ (where $|\overline{\Omega}_b|= |\Omega_b|$).   The $K_aK_b$ vectors
\begin{equation}
{\bf p}_a^\gamma \otimes {\bf p}_b^\delta ~~~\text{for}~~ \gamma\delta\in \overline{\Omega}_a\times\overline{\Omega}_b
\end{equation}
are linearly independent and, similarly, the $K_aK_b$ vectors
\begin{equation}
{\bf r}_a^\alpha\otimes {\bf r}_b^\beta ~~~\text{for}~~ \alpha\beta\in \Omega_a\times\Omega_b
\end{equation}
are linearly independent.   It follows from (\ref{fidcompeqn}) and some simple linear algebra that the $K_aK_b$ vectors
\begin{equation}\label{abstates}
{\bf p}_{ab}^{\gamma\delta} ~~~\text{for}~~ \gamma\delta\in \overline{\Omega}_a\times\overline{\Omega}_b
\end{equation}
are linearly independent as are the $K_aK_b$ vectors
\begin{equation}\label{locrs}
{\bf r}_{ab}^{\alpha\beta}  ~~~\text{for}~~ \alpha\beta\in \Omega_a\times\Omega_b
\end{equation}
From this we can prove the following
\begin{quote}
{\bf Theorem 4}~ For composite systems
we can choose $\Omega_{ab}$ such that
\begin{equation}\label{omebaabinab}
\Omega_a\times\Omega_b \subseteq \Omega_{ab}
\end{equation}
\end{quote}
This theorem has an immediate corollary:
\begin{quote}
{\bf Corollary 2}~ For composite systems $K_{ab} \geq K_a K_b$
\end{quote}
Here $K_a = |\Omega_a|$.  This inequality follows from fact that we have at least $K_aK_b$ linearly independent states in (\ref{abstates}).   The set relation (\ref{omebaabinab}) follows since the effects in (\ref{locrs}) are linearly independent and hence that we can choose $K_aK_b$ of the $K_{ab}$ fiducial effects in $\Omega_{ab}$ to correspond to local effects.  By a local effect we mean one comprised of disjoint circuit fragments, one on system $a$ and one on system $b$.  It follows from (\ref{omebaabinab}) that we can write
\begin{equation}
\Omega_{ab}=\breve\Omega_{ab} \cup \tilde\Omega_{ab} ~~~\text{where}~~ \breve\Omega_{ab}=\Omega_a\times\Omega_b
\end{equation}
The fiducial effects in $\breve\Omega_{ab}$ are local.
Hence, we can write a general bipartite state, with preparation $\varepsilon$, as
\begin{equation}\label{locnonloc}
{\bf p}_{ab}^{\varepsilon} = \breve{\bf p}_{ab}^{\varepsilon} \oplus \tilde{\bf p}_{ab}^{\varepsilon}
\end{equation}
where the elements of $\breve{\bf p}_{ab}^{\varepsilon}$ are the probabilities corresponding to effects in $\breve{\Omega}_{ab}$ and the elements of $\tilde{\bf p}_{ab}^{\varepsilon}$ are the probabilities corresponding to the effects in $\tilde{\Omega}_{ab}$.
Note that $\breve{\bf p}_{ab}^{\varepsilon}$ lives in the tensor product space of the vector spaces for component systems because $\breve\Omega_{ab}=\Omega_a\times\Omega_b$.  If systems $1$ and $2$ are both in homogeneous states then it follows from equation (\ref{fidcompeqn}) that
\begin{equation}\label{pureotimespure}
\breve{\bf p}_{ab}^{\gamma\delta} = \nu_{\gamma\delta}{\bf p}_a^{\gamma}\otimes {\bf p}_b^{\delta}
\end{equation}
A similar result holds even if only one system is in a homogeneous state (this follows from theorem 1).
These results for bipartite systems generalize in the obvious way to composites having more than two component systems (for three systems we use  $\Omega_{abc}=(\Omega_a\times\Omega_b\times\Omega_c) \cup \tilde \Omega_{abc}$).

It is easy to see that if each system is subject to its own local transformation (so the circuit fragments corresponding to the transformations are disjoint) then the state updates as
\begin{equation}\label{locnonlocZZ}
{\bf p}_{cd}^{\varepsilon\alpha\beta} = [({}^c\!Z_a^\alpha\otimes {}^d\! Z_b^{\beta})\breve{\bf p}_{ab}^{\varepsilon}] \oplus {}^{cd}\!\tilde{Z}_{ab}^{\alpha\beta}{\bf p}_{ab}^{\varepsilon}
\end{equation}
where the form of the ${}^{cd}\!\tilde{Z}_{ab}^{\alpha\beta}$ matrix depends, in general, on the particular theory (it acts on ${\bf p}_{ab}^\varepsilon$ to give $\tilde{\bf p}_{cd}^{\varepsilon\alpha\beta}$).  This equation follows by considering the case where both systems are in homogeneous states.  Then we have $K_aK_b$ linearly independent vectors $\breve{\bf p}_{ab}^{\gamma\delta} = \mu_{\gamma\delta}{\bf p}_a^{\gamma}\otimes {\bf p}_b^{\delta}$ with ($\gamma\delta\in\overline{\Omega}_a\times\overline{\Omega}_b$).  The $\breve{\bf p}$ part of the state must remain as a tensor product like this after the local transformations to ensure consistency with equation (\ref{fidcompeqn}) (notice that if it did not then there would exist a correlation-revealing measurement contradicting Theorem 2). But the vectors ${\bf p}_a^{\gamma}\otimes {\bf p}_a^{\delta}$ span the space of possible $\breve{\bf p}_a$ vectors and so (\ref{locnonlocZZ}) must be true generally.

If system $d$ is the null system (so the transformation on system $2$ is an effect) then the $\tilde Z$ matrix has no elements and we can write
\begin{equation}\label{locnonlocZnull}
{\bf p}_{c}^{\varepsilon\alpha\beta} = ({}^c\!Z_a^\alpha\otimes {}^-\!Z_b^{\beta})\breve{\bf p}_{ab}^{\varepsilon}
\end{equation}
(Note that if the $\tilde Z$ matrix did have elements then the two sides of this equation would be column vectors of different lengths.)  In particular, the reduced state of system $1$ is given by
\begin{equation}\label{reducedstate}
{\bf p}_{a}^\varepsilon\equiv{\bf p}_{a}^{\varepsilon 0 -} = [{}^a\!Z_a^0\otimes ({\bf r}_b^-)^T]\breve{\bf p}_{ab}^{\varepsilon}
\end{equation}
(recall that ${}^a\!Z_a^0$ is the identity).  Hence the reduced state at either end depends only on the elements in the $\breve{\bf p}_{ab}^\varepsilon$ part of ${\bf p}_{ab}^\varepsilon$.

If both systems $c$ and $d$ are null then (\ref{locnonlocZnull}) becomes
\begin{equation}\label{localmeasurements}
p^{\varepsilon\alpha\beta} = ({\bf r}_a^\alpha\otimes{\bf r}_b^\beta)\cdot \breve{\bf p}_{ab}^\varepsilon
\end{equation}
This also follows directly from (\ref{fidcompeqn}) and the fact that the vectors ${\bf p}_a^{\gamma}\otimes {\bf p}_a^{\delta}$ span the space of possible $\breve{\bf p}_a$ vectors.  This equation tells us that all local effects are linear combinations of the fiducial effects corresponding to the $\breve{\Omega}_{ab}$ part of $\Omega_{ab}$.  Hence, all effects ${\bf r}_{ab}^{\gamma}$ with $\gamma\in \tilde\Omega_{ab}$ are nonlocal - the corresponding circuit fragments cannot consist of disjoint parts acting separately on $a$ and $b$. Theories in which the state can be entirely determined by local measurements are called {\it locally tomographic}.  This gives us an important theorem
\begin{quote}
{\bf Theorem 5}~ Theories having $K_{ab}=K_aK_b$ are locally tomographic and vice-versa.
\end{quote}
This corresponds to the case where $\tilde\Omega_{ab}$ is the null set.

\subsection{Homogeneity and uncorrelatability are equivalent notions}

We define:
\begin{quote}
{\bf An uncorrelatable state} is one having the property that a system in this state cannot be correlated with any other system (so that any joint probabilities factorize).
\end{quote}
Let ${\bf p}_a^\gamma $ be an uncorrelatable state.  Let ${\bf p}^{\gamma}_{ab}$ be a state for a bipartite system having the  property that its reduced state is ${\bf p}_a^\gamma$.   If the bipartite system is prepared in any such state then the joint probabilities will, by definition, factorize.  We will prove
\begin{quote}
{\bf Theorem 6}~ If we allow arbitrary mixtures then all homogeneous states are uncorrelatable and all uncorrelatable states are homogeneous.
\end{quote}
That is homogeneity and uncorrelatability are equivalent notions.  It follows immediately from Theorem 2 that homogeneous states are uncorrelatable.  To prove that uncorrelatable states are homogeneous we assume the converse.  Thus assume that the heterogeneous state ${\bf p}_a^\gamma = {\bf p}_a^{\gamma_1}+ {\bf p}_a^{\gamma_2}$ (where the two terms are non-parallel) is uncorrelatable.  Since the state is assumed to be uncorrelatable we must be able to write the state of the composite system as
\begin{equation}
{\bf p}_{ab}^{\gamma} =  \mu_{\gamma}[{\bf p}_a^{\gamma}\otimes {\bf p}_b^{\gamma}] \oplus \tilde{\bf p}_{ab}^{\gamma}
\end{equation}
since otherwise the probability does not factorize for all $\alpha\beta\in\Omega_a\times\Omega_b$.  Hence
\begin{equation}\label{contradiction}
{\bf p}_{ab}^{\gamma} = \mu_{\gamma} [{\bf p}_a^{\gamma_1}\otimes {\bf p}_b^{\gamma}] \oplus \tilde{\bf p}_{ab}^{\gamma} +
\mu_{\gamma} [{\bf p}_a^{\gamma_2}\otimes {\bf p}_b^{\gamma}] \oplus \tilde{\bf p}_{ab}^{\gamma}
\end{equation}
But, using (\ref{reducedstate}), we see that the reduced state, ${\bf p}_a^\gamma$, for system $1$ of (\ref{contradiction}) is parallel to the reduced state, ${\bf p}_a^{\gamma'}$, of
\begin{equation}\label{correlatedstate}
{\bf p}_{ab}^{\gamma'} = \mu_{\gamma_1\delta_1}[{\bf p}_a^{\gamma_1}\otimes {\bf p}_b^{\delta_1}] \oplus \tilde{\bf p}_{ab}^{\gamma_1\delta_1} +
\mu_{\gamma_2\delta_2} [{\bf p}_a^{\gamma_2}\otimes {\bf p}_b^{\delta_2}] \oplus \tilde{\bf p}_{ab}^{\gamma_2\delta_2}
\end{equation}
where we choose any two distinct ${\bf p}_b^{\delta_1}$ and ${\bf p}_b^{\delta_2}$ having normalization such that $\mu_{\gamma_1\delta_1}=\mu_{\gamma_2\delta_2}$.  It is possible to choose two distinct states like this if there exist systems that are non-trivial in the sense that they require more than one fiducial effect. (If all systems are trivial then all states are homogeneous and so, by Theorem 2, all states are uncorrelatable in any case.) We allow arbitrary mixtures and so can take mixtures with the null state to make sure ${\bf p}_b^{\delta_1}$ and ${\bf p}_b^{\delta_2}$ have normalization so that $\mu_{\gamma_1\delta_1}=\mu_{\gamma_2\delta_2}$.   The state (\ref{correlatedstate}) is preparable by taking a mixture of the preparations $\gamma_1\delta_1$ and $\gamma_2\delta_2$.  This state is clearly correlated.   By taking a mixture with the null state for the longer of ${\bf p}_a^\gamma$ and ${\bf p}_a^{\gamma'}$ we obtain two equal states one of which is uncorrelatable by assumption and one of which is correlatable by the above proof.  Hence our assumption was false and it follows that uncorrelatable states are homogeneous.

\subsection{Probabilities for disjoint circuit}\label{disjointfrags}

If we have two disjoint setting-outcome specified circuits, $\alpha$ and $\beta$, then expect the joint probability to factorize
\begin{equation}\label{alphabetafactorise}
p^{\alpha\beta} = p^{\alpha}p^\beta
\end{equation}
A simple application of Bayes' rule shows that (\ref{alphabetafactorise}) is equivalent to demanding that the probability associated with a circuit is independent of the outcomes seen at other disjoint circuits.  This is an extremely natural condition since otherwise we would have to take into account all the outcomes seen on all other disjoint circuits in the past which form a part of our memory before writing down a probability for the circuit.  On the other hand, one can easily envisage a situation in which the probability is not independent of outcomes elsewhere.  For example, the eventual outcome of a spinning coin might be correlated with the outcome of an apparently disjoint experiment which is, incidently, influenced by photons scattered from the coin while it spins.  More generally, if there are hidden variables, then there may be correlations between outcomes even though the marginals are independent of what happens at the other side.  It is not clear that disjointness of the circuits is enough to prevent such correlations.  In view of this, it is interesting that the following theorem holds.
\begin{quote}
{\bf Theorem 7}~ If there exists at least one type of system which can be prepared in a pure state of norm one then the probability associated with any circuit is independent of the settings on any other disjoint circuit.
\end{quote}
Let the preparation associated with this pure state of norm one be $\gamma$.  Consider the circuit $(\gamma -)(\gamma -) (\alpha)(\beta)$ consisting of two instances of the circuit obtained by performing the trace effect on a preparation $\gamma$, and two more disjoint circuits $\alpha$ and $\beta$.  Thus we have four disjoint circuits in total.  We can regard this circuit as consisting of the effect $-\alpha$ on one of the $\gamma$ preparations and the effect $-\beta$ on the other $\gamma$ preparation.  Then we have
\begin{equation}
p^{(\gamma-\alpha)(\gamma-\beta)} p^{(\gamma-)(\gamma-)} = p^{\gamma-\alpha}p^{\gamma-\beta}
\end{equation}
by Theorem 2 (equation (\ref{gammafactorise}) in particular).  But $p^{\gamma-}=1$ since the state is of norm one.  Hence, using Bayes' rule, (\ref{alphabetafactorise}) follows and the theorem is proved.

We say that a set of operations is {\it fully closable} if it is closable and if the probability for a circuit is independent of the outcomes seen at other disjoint circuits.  It follows from Theorem 7 that closable sets of operations admitting at least one pure state of norm one are fully closable.  In the case that we have a fully closable set of operations it is clear that we can write the $\breve{\bf p}_{ab}$ part of the state associated with disjoint preparations as
\begin{equation}
\breve{\bf p}_{ab}^{\gamma\delta} = {\bf p}_a^{\gamma}\otimes {\bf p}_b^{\delta}
\end{equation}
It is interesting to note that (\ref{pureotimespure}) is an example of this with $\nu_{\gamma\delta}=1$ which clearly follows from (\ref{strange}) when probabilities for disjoint circuits factorize.





\subsection{Examples of the relationship between $K_{ab}$, $K_a$, and $K_b$}

If $N_a$ is the number of states that can be distinguished in a single shot measurement then it is reasonable to suppose $N_{ab}=N_aN_b$. This is true in all the examples we will discuss. In classical probability theory $K_a=N_a$.  In quantum theory $K_a=N_a^2$.  Hence $K_{ab}=K_aK_b$ and so, by Theorem 5, we have local tomography in these theories. In real Hilbert space quantum theory, where the state is represented by a positive density matrix with real entries, we have $K_a= N_a+ N(N-1)/{2!}$.  This has $K_{ab}> K_aK_b$ which is consistent with Corollary 2.  However, quaternionic quantum theory has $K_a= N_a+ 4N(N-1)/{2!}$ which has $K_{ab} < K_aK_b$. This is inconsistent with Corollary 2 and hence quaternionic quantum theory cannot be formulated in this framework.  Since we have made very minimal assumptions (only that we have closable sets of operations in an operational framework) it seems that quaternionic quantum theory is simply an inconsistent theory (at least for the finite $K_a$ case considered here).

\section{Theories for which $K_{ab}=K_aK_b$}

\subsection{Motivation for local tomography}

Of the examples we just considered, the two corresponding to real physics are both locally tomographic having $K_{ab}=K_aK_b$.  This is a very natural property for a theory to have (it is one of the axioms in \cite{Hardy1}).  It says that, from a counting point of view, no new properties come into existence when we put two systems together.  It allows a certain very natural type of locality so that it is possible to characterize a system made from many parts by looking at the components.  It implies that the full set of states for a composite system requires the same number of parameters for its specification as the separable states (formed by taking mixtures of states prepared by disjoint circuits).  Given that this is such a natural constraint we will study it a little more closely.  We will also give axioms for classical probability and quantum theory since they are examples of this sort.

\subsubsection{Operation locality}\label{locality}

An extremely useful property of locally tomographic theories is that they are local in the sense that the state is updated by the action of local matrices at each operation.  We will call this property {\it operation locality}. We see from (\ref{locnonloc}) that if $\tilde\Omega_{ab}=\emptyset$ then $\tilde{\bf p}_{ab}^\gamma=0$ and ${\bf p}_{ab}^\gamma=\breve{\bf p}_{ab}^\gamma$ and hence according to (\ref{locnonlocZZ}) we see that under local transformations at each end (corresponding to disjoint circuit fragments) the state will update as
\begin{equation}
{\bf p}_{cd}^{\gamma\alpha\beta} = ({}^c\!Z_a^\alpha\otimes {}^d\! Z_b^{\beta}){\bf p}_{ab}^{\gamma}
\end{equation}
Hence
\begin{equation}\label{ZotimesZ}
{}^{cd}\! Z_{ab}^{\alpha\beta} = {}^c\!Z_a^\alpha\otimes {}^d\! Z_b^{\beta}
\end{equation}
for transformations corresponding to disjoint circuit fragments.  In particular this implies
\begin{equation}\label{withidentitytrue}
{}^{cb}\!Z_{ab}^{\alpha 0} = {}^{c}\!Z_{a}^{\alpha} \otimes {}^{b}\!Z_{b}^{0}
\end{equation}
This is equation (\ref{withidentity}) we speculated about in Section \ref{commutationsection}. This means an operation has a trivial effect on systems that do not pass through it.  If we have a fully closable set of transformations (as long as there exists a least one state of norm-one this follows from Theorem 7) then we can specialize this equation to the case of null input states (where $a=-$ and/or $b=-$) since the state prepared by disjoint preparations is a product state.  We will assume this in what follows.

\begin{figure*}[t]
\centering
{\includegraphics[width=0.4\textwidth]{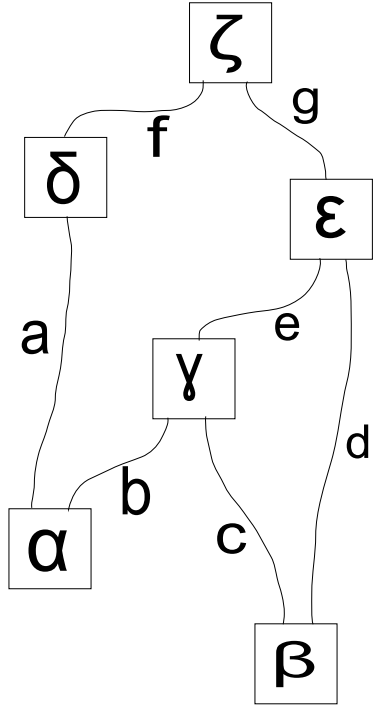}} \caption{\small We show how to calculate the probability associated with this simple example in the text using a $Z$ matrix for each operation.}\label{simpleex}
\end{figure*}

The great thing about (\ref{ZotimesZ}) is that it can be used to calculate the probability for any circuit using a $Z$ matrix for each operation.   To do this we choose a complete foliation and then use the tensor product to combine operations at each time step.  One way of calculating the probability $p^{\alpha\beta\gamma\delta\varepsilon\zeta}$for the example shown in Fig.\ \ref{simpleex} is
\begin{equation}
Z^\zeta_{fg}(\,{}^f\!Z_f^0\otimes \,{}^g\!Z_{ed}^\varepsilon)
(\,{}^f \! Z_f^0\otimes \,{}^e \!Z_{bc}^\gamma \otimes \,{}^d \!Z_d^0)
( \,{}^f\! Z_a^\delta \otimes \,{}^b \! Z_b^0 \otimes \,{}^c\! Z_c^0 \otimes \,{}^d\! Z_d^0)
( \,{}^{ab}\! Z^\alpha\otimes \,{}^{cd}\! Z^\beta)
\end{equation}
While it is very satisfying that the calculation can be broken down like this, it is unfortunate that we have to pad out the calculation with lots of identity matrices like ${}^c\! Z_c^0$.  This means that there are more matrices than operations in this calculation.  Relatedly, we have to be very careful what order we take the product of all these matrices (it has to correspond to some complete foliation).  In the causaloid approach \cite{Hardy2, Hardy3, Hardypirsa1} we will have neither of these problems.  We simply take what is called the causaloid product of a vector associated with each operation without regard for the order and without having to pad out the calculation with identity matrices.

\subsection{Classical probability theory}

It is very easy to characterize classical probability theory in this framework. It is fully characterized by the following two axioms:
\begin{description}
\item[Composition] $K_{ab}=K_aK_b$
\item[Transformations] Transformation matrices, $ {}^c\! Z_a^\alpha$, have the property that the entries are nonnegative and the sum of the entries in each column is less than or equal to $1$.
\end{description}
To see this is equivalent to usual presentations of classical probability theory note the following.  We can interpret $K_a\equiv N_a$ as the maximum number of distinguishable states for this classical system (for a coin we have $N_a=2$, for a die $N_b=6$).  The state is given by a ${\bf p}_a^\alpha = \,{}^a\! Z^\alpha$ and is a column vector.  The sum of the entries in this vector must be less than or equal to 1. They can be interpreted as the probabilities associated with each of the distinguishable outcomes (for a die they are the probabilities associated with each face).  The trace effect is given by $({\bf r}_a^-)= Z_a^-$. It is a row vector.  The value of each entry is 1 and this is consistent with the constraint that the sum of the columns cannot be greater than 1.  Norm preserving transformations are stochastic matrices.  Since $K_{ab}=K_aK_b$, we have the operation locality property and so we can calculate the probability for an arbitrary circuit from matrices for the operations that comprise it.


\subsection{Quantum theory}

To give the rules for Quantum Theory we need a few definitions first.  Let ${\mathcal H}_{N_a}$ be a complex Hilbert space of dimension $N_a$.  Let ${\mathcal V}_{N_a}$ be the space of Hermitian operators that act on this.  All positive operators
are Hermitian.  Furthermore, it is possible to find a set of $N_a^2$ linearly independent positive operators that span ${\mathcal V}_{N_a}$.  Let $\widehat P_a^\alpha$ for $\alpha\in \Omega_a$ be one such set.  Define
\begin{equation}
\widehat{\bf P}_a = \left(\begin{array}{c} \vdots \\ \widehat{P}_a^\alpha \\ \vdots \end{array} \right) ~~~~ \alpha\in \Omega_a
\end{equation}
A {\it positive map} ${}^c\$_a $ from ${\mathcal V}_{N_a}$ to ${\mathcal V}_{N_c}$ is one which acts on a positive operator $\rho_a\in {\mathcal V}_{N_a}$ and returns a positive operator $\rho'_c\in {\mathcal V}_{N_c}$ for any positive operator $\rho_a$.  The map ${}^c\$_a $ is {\it completely positive} if $ {}^c\$_a \otimes \,{}^b\!I_b$ is a positive map from ${\mathcal V}_{N_aN_b}$ to ${\mathcal V}_{N_cN_b}$for any $b$ where ${}^b\!I_b$ is the identity map on ${\mathcal V}_{N_b}$.  Further, we want our maps to have the property that they do not lead to probabilities greater than 1.  We will demand that they must be completely trace non-increasing when they act on density matrices. This means that $ {}^c\$_a \otimes \,{}^b\!I_b$ must be trace non-increasing for any $b$.  Quantum theory is fully characterized by the following two axioms.
\begin{description}
\item[Composition] $K_{ab}=K_aK_b$
\item[Transformations] Transformation matrices are of the form
\begin{equation}\label{quantumtrans}
{}^c\! Z_a^\alpha = \text{Trace}\big(\widehat{\bf P}_c \,{}^c\$_a^\alpha(\widehat{\bf P}_a^T)\big)
\big[\text{Trace}( \widehat{\bf P}_a  \widehat{\bf P}_a^T)  \big]^{-1}
\end{equation}
where ${}^c\$_a^\alpha$ is completely positive and completely trace non-increasing and $T$ denotes transpose.
\end{description}
This is a much more compact statement of the rules of quantum theory than is usually given.  We will make a few remarks to decompress this.   First note that $\text{Trace}\big(\widehat{\bf P}_c \,{}^c\$_a^\alpha(\widehat{\bf P}_a^T)\big)$ is a $K_c\times K_a$ matrix having $\beta\gamma$ element $\text{Trace}\big(\widehat{P}_c^\beta \,{}^c\$_a^\alpha(\widehat{P}_a^\gamma)\big)$ with $\beta\in\Omega_c$ and $\gamma\in\Omega_a$. By defining
\begin{equation}
{\bf p}_a^\delta = \text{Trace}(\widehat{\bf P}_a \rho_a^\delta)
\end{equation}
where $\rho_a^\delta$ is the usual quantum state, and using $\widehat \rho^\alpha_a=\widehat{\bf P}_a\cdot {\bf s}^\alpha$ (since $\widehat\rho_a$ must be given by some sum of the linearly independent spanning set), it can be shown after a few lines of algebra that
\begin{equation}
{\bf p}_c^{\delta\alpha} = {}^c\! Z_a^\alpha {\bf p}_a^{\delta} ~~~\Leftrightarrow ~~~
\widehat{\rho}_a^{\,\delta\alpha} ={}^c \$_a^\alpha (\widehat{\rho}_a^\delta)
\end{equation}
Hence we get the correct transformations with (\ref{quantumtrans}).  Note that, as in the classical case, we can write a state as ${\bf p}_a^\alpha = \,{}^a\! Z^\alpha$.  This state is associated with a completely positive map ${}^a \$^\alpha$ where the absence of an input label implies that we have a null input system which corresponds to a one dimensional Hilbert space. This must have trace less than or equal to one since otherwise $ {}^a\$ \otimes \,{}^b\!I_b$ would be trace increasing (this is the reason we impose that the map should be completely trace non-increasing rather than just trace non-increasing).  Also note that for ${}^a\$_a^0$ (the identity map) we get the identity for ${}^a\! Z_a^0$ in (\ref{quantumtrans}) as we must.  The composition rule $K_{ab}=K_aK_b$ implies that $N_{ab}=N_aN_b$ (since we see from inspection of the rank of the matrices that $|\Omega_a|=N_a^2$) and hence the tensor product structure for Hilbert spaces corresponds to the tensor product structure discussed in Sec.\ \ref{locality}.  The fact that we have $K_{ab}=K_aK_b$ means that we have the operation locality property.  We can calculate the probability for a general circuit using these $Z$ matrices for each of the operations.  Hence our list of postulates for Quantum Theory is complete.

\subsection{Reasonable postulates for quantum theory}

The objective of this chapter has been to set up a general probabilistic framework. It is worth mentioning that we can give the following very reasonable postulates which enable us to reconstruct quantum theory within this framework.
\begin{description}
\item[Information] Systems having, or constrained to have, a given information carrying capacity have the same properties.
\item[Composites] Information carrying capacity is additive and local tomography is possible (i.e. $N_{ab}=N_aN_b$ and $K_{ab}=K_aK_b$).
\item[Continuity] There exists a continuous reversible transformation between any pair of pure states.
\item[Simplicity] Systems are described by the smallest number of probabilities consistent with the other postulates.
\end{description}
We can show from the first two postulates that $K=N^r$ where $r=1, 2, \dots$.  The continuity postulate rules out the classical probability case where $K=N$.  The simplicity postulate then implies that we have $K=N^2$.  We construct the Bloch sphere for the $N=2$ case using, in particular, the continuity postulate.  Then the information postulate and composites postulate are used to obtain quantum theory for general $N$. We refer the reader to \cite{Hardy1, Hardypirsa2} for details.

\section{Conclusions}

We have exhibited a very natural framework for general probabilistic theories in an operational setting.  We represent experiments by circuits and have been particularly careful to give operational interpretations to the elements of these circuits (operations are single uses of an apparatus and wires represent apertures being placed next to one another). By considering closable sets of operations we are able to introduce probabilities and then to set up the full theory wherein $Z$ matrices are associated with circuit fragments.  The special case of locally tomographic theories has the operation locality property so that we can combine $Z$ matrices corresponding to circuit fragments that are in parallel using the tensor product.  This enables us to break down a calculation into smaller parts.  The framework here is still lacking.  Most crucially, it is only able to take into account one particular way in which operations can be connected (corresponding to placing apertures next to each other) but there are many other ways.  A more general theory is under development to allow more for other types of connections (see \cite{Hardypirsa1}).  The framework is discrete and hence is not readily adaptable to quantum field theory.  It would be very interesting either to develop a continuous version or to show how quantum field theory can be fully understood in such a discrete framework.  Algebraic quantum field theory can be understood in operational terms (see for example Haag \cite{Haag}).  However, putting the issue of discreteness aside, it is a rather less general operational theory than that presented in this chapter and so there may be advantages to studying quantum field theory in the framework presented here.  It is worth saying that there is a tension between operationalism and use of the continuum in physics.  From an operational point of view, the continuum is best understood as a mathematical tool enabling us to talk about a series of ever more precise experiments.  It is possible that such a series may, eventually, be better described with other mathematical tools.

There are two types of motivation for considering general probabilistic theories.  First, we may be able to better formulate and understand our present theories within these frameworks.  It may be possible to write down a set of postulates or axioms which can be used to reconstruct these theories within such a framework.  For the case of quantum theory there has been considerable work of this nature already.  It would be interesting to see something similar for general relativity.  In particular, there ought to be a simple and elegant formulation of general relativity for the case where there is probabilistic ignorance of the value of quantities that might be measured in general relativity,  (let us call this probabilistic general relativity).  Such a theory might be best understood in a general probabilistic framework (though probably more general than the one presented in this chapter) \cite{Hardypirsa1}.   The second reason to consider general probabilistic theories is to try to go beyond our present theories.  The most obvious application would be to work towards a theory of quantum gravity (see \cite{Hardy2, Hardy3}).  The program of constructing general probabilistic theories and then constraining then using some principles or postulates may free us from the hidden mathematical obstacles to formulating quantum gravity that stand in the way of the more standard approaches such as string theory and loop quantum gravity.

\section*{Acknowledgements}
Research at Perimeter Institute for Theoretical Physics is supported in part by
the Government of Canada through NSERC and by the Province of Ontario through MRI.   I am grateful to Vanessa Hardy for help with the manuscript.




\begin{thebibliography}{00}
\bibitem{Einstein1905} A. Einstein. Zur Elektrodynamik bewegter K\"orper.  \textit{Annalen der Physik}, \textbf{17} (1905) 891-921.
\bibitem{Fuchslittlemore} C.A. Fuchs. Quantum mechanics as quantum information (and only a little more).  arXiv:quant-ph/0205039 (2002).  http://arxiv.org/abs/quant-ph/0205039.
\bibitem{RQTpirsa} Conference: Reconstructing quantum theory, organised by Philip Goyal and Lucien Hardy. PIRSA:C09016 (2009). http://pirsa.org/c09016.
\bibitem{Hardy2} L. Hardy. Probability theories with dynamic causal structure: A new framework for quantum gravity.  arXiv:gr-qc/0509120 (2005). http://arxiv.org/abs/gr-qc/0509120.
\bibitem{Hardypirsa1} L. Hardy. Operational structures as a foundation for probabilistic theories. PIRSA:09060015 (2009). http://pirsa.org/09060015.
\bibitem{Hardy1} L. Hardy. Quantum theory from five reasonable axioms.  arXiv:quant-ph/0101012 (2001). http://arxiv.org/abs/quant-ph/0101012.
\bibitem{Hardy3} L. Hardy. Towards quantum gravity: A framework for probabilistic theories with non-fixed causal structure. \textit{Journal of Physics}, \textbf{A40} (2007) 3081-3099.
\bibitem{Hardypirsa2} L. Hardy. Operational structures and natural postulates for quantum theory. PIRSA:09080011 (2009). http://pirsa.org/09080011.
\bibitem{causalset} R. Sorkin. Spacetime and causal sets. In J. D'Olivo et. al., editor, Relativity and Gravitation:
Classical and Quantum. World Scientific, (1991).
\bibitem{ch1} R.B. Griffiths. Consistent histories and the interpretation of quantum mechanics. \textit{Journal of Statistical Physics}, \textbf{36} (1984) 219-272.
\bibitem{ch2}M. Gell-Mann and J.B. Hartle. Classical equations for quantum systems. \textit{Physical Review
D}, \textbf{47} (1993) 3345–3382.
\bibitem{ch3} R. Omn\`es. The Interpretation of quantum mechanics. Princeton Univ. Press, 1994.
\bibitem{ch4}  J.B. Hartle. Spacetime quantum mechanics and the quantum mechanics of spacetime. \textit{Gravitation and Quantizations: Proceedings of the 1992 Les Houches Summer School}, ed. by B. Julia and J. Zinn-Justin, North Holland, Amsterdam, (1995).
\bibitem{Markopoulou} F. Markopoulou. Quantum causal histories. \textit{Classical and Quantum Gravity}, \textbf{17} (2000) 2059–2077.
\bibitem{Prakash} R.F. Blute, I.T. Ivanov, P. Panangaden. Disgrete quantum causal dynamics.  \textit{International Journal of Theoretical Physics} \textbf{42} (2003) 2025-2041.
\bibitem{Prakashpirsa} P. Panangadan.  Discrete quantum causal dynamics. PIRSA:09060029 (2009).  http://pirsa.org/09060029/.
\bibitem{Leifer} M. Leifer. Quantum causal networks.  PIRSA:06060063 (2006). http://pirsa.org/06060063.
\bibitem{AbramskyCoecke} S. Abramsky and B. Coecke. A categorical semantics of quantum protocols.
\textit{Proceedings of the 19th Annual IEEE Symposium on Logic in Computer Science
(LICS '04)}, (2004) 415–425.
\bibitem{Coeckeotherstuff} B. Coecke. Where quantum meets logic, . . . in a world of pictures! PIRSA:09040001 http://pirsa.org/09040001
\bibitem{Mackey} G. Mackey. \textit{Mathematical Foundations of Quantum Mechanics}. Benjamin (1963).
\bibitem{Ludwig} G. Ludwig, \textit{An Axiomatic Basis of Quantum Mechanics} volumes 1 and 2. Springer-Verlag (1985, 1987).
\bibitem{DaviesLewis} E. B. Davies and J. T. Lewis, An operational approach
to quantum probability, \textit{Communications of Mathematical Physics} \textbf{17} (1970) 239-260.
\bibitem{Araki} H. Araki. On a characterization of the state space of quantum mechanics. \textit{Communications of Mathematical Physics} \textbf{75} (1980) 1–24.
\bibitem{Gudder} S. Gudder, S. Pulmannov\'a, S. Bugajski, and
E. Beltrametti. Convex and linear effect algebras.
\textit{Reports on Mathematical Physics} \textbf{44} (1999) 359-379.
\bibitem{FoulisRandall} D. J. Foulis and C. H. Randall, Empirical logic and tensor products, in H. Neumann, (ed.), \textit{Interpretations and Foundations of Quantum Theory}, Bibliographisches Institut, Wissenschaftsverlag, Mannheim (1981).
\bibitem{Barrett} J. Barrett. Information processing in generalized probabilistic theories. \textit{Physical
Review A} \textbf{75} (2007) 032304.
\bibitem{BBLW1} H. Barnum, J. Barrett, M. Leifer, and A. Wilce. Cloning and broadcasting in
generic probabilistic models. arXiv.org:quant-ph/0611295, (2006). http://arxiv.org/abs/0611295.
\bibitem{BBLW2} H. Barnum, J. Barrett, M. Leifer and A.Wilce, A general
no-cloning theorem, Phys. Rev. Lett. 99 240501 (2007).
\bibitem{BBLW3} H. Barnum, J. Barrett, M. Leifer, and A. Wilce. Teleportation in general
probabilistic theories. arXiv.org:0805.3553, (2008). http://arxiv.org/abs/0805.3553.
\bibitem{BW1}  H. Barnum and A. Wilce. Information processing in convex operational theories. Proceedings
of QPLV-DCMIV, Electronic Notes in Theoretical Computer Science (2008).
\bibitem{BW2}  H. Barnum and A. Wilce. Ordered linear spaces and categories as frameworks for information-processing characterizations of quantum and classical theory. arXiv:0908.2354 (2009).  http://arxiv.org/abs/0908.2354.
\bibitem{Wootters} W.K. Wootters.  Local accessibility of quantum states, in Complexity, entropy and the physics
of information edited by W. H. Zurek (Addison-Wesley, 1990)
\bibitem{Wootters1} W.K. Wootters, Quantum mechanics without probability amplitudes.  \textit{Foundations of Physics} \textbf{16} (1986) 391.
\bibitem{PRpaper}  S. Popescu and D. Rohrlich. Nonlocality as an axiom. \text{Foundations of Physics} \textbf{24} (1994) 379.
\bibitem{Pawlowski} M. Pawlowski, T. Paterek, D. Kazlikowski, V. Scarani, A. Winter and M. \.Zukowski, et al. A new physical principle: Information causality. \textit{Nature} \textbf{461} (2009) 1101.
\bibitem{Gross} D. Gross, M. Mueller, R. Colbeck, O.C.O. Dahlsten. All reversible dynamics in maximally non-local theories are trivial. arXiv:0910.1840v1 (2009). http://arxiv.org/abs/0910.1840.
\bibitem{DAriano} G.M. D'Ariano. How to Derive the Hilbert-Space formulation of quantum mechanics from purely operational axioms. arXiv:quant-ph/0603011 (2006). http://arxiv.org/abs/quant-ph/0603011.
\bibitem{CDP} G. Chiribella, G.M. D'Ariano, P. Perinotti. Probabilistic theories with purification. arXiv:0908.1583 (2009). http://arxiv.org/abs/0908.1583.
\bibitem{Gillies} D. Gillies, \textit{Philosophical theories of probability}, Routledge (2000).
\bibitem{Haag} R. Haag. \textit{Local quantum physics: Fields, particles, algebras.} Springer (1992).
\end{thebibliography}
\end{document}